\documentclass[letterpaper,titlepage,11pt]{article}
\usepackage{hyperref}
\usepackage{amssymb,amsmath,amsfonts}
\usepackage{epsfig}

\def\clock{{\count0=\time
           \divide\count0 60
           \ifnum\count0<10 0\fi\the\count0
           \multiply\count0 -60 \advance\count0 \time
           :\ifnum\count0<10 0\fi \the\count0
         }}
\newcommand{\timestamp}{{\small\vbox{\hbox{\tt\jobname.tex}
\hbox{\the\day/\the\month/\the\year, \clock}}}}

\newcommand{\ds}{\displaystyle}
\newcommand{\be}{\begin{equation}} \newcommand{\ee}{\end{equation}}
\newcommand{\bea}{\begin{eqnarray}} \newcommand{\eea}{\end{eqnarray}}

\newcommand{\CO}{\mathcal{O}}

\newcommand{\CM}{\mathcal{M}}

\newcommand{\id}{\hbox{1\kern-.27em l}}
\newcommand{\sid}{\hbox{\scriptsize1\kern-.27em l}}

\newcommand{\we}{\kern-.1em\wedge\kern-.1em}
\newcommand{\scal}{\kern-.13em\cdot\kern-.13em}

\newcommand{\II}{I\kern-.09em I}

\newcommand{\al}{\alpha} \newcommand{\ga}{\gamma}
 \newcommand{\bet}{\beta}
\newcommand{\ka}{\kappa} 
 
\newcommand{\la}{\lambda} 
 \newcommand{\Om}{\Omega}

\newcommand{\R}{\mathbb{R}}

\newcommand{\spa}{\ , \ \ }

\newcommand{\eps}{\epsilon}

\newcommand{\beastar}{\begin{eqnarray*}}
\newcommand{\eeastar}{\end{eqnarray*}}

\newcommand{\mt}{\mathfrak{t}}
\newcommand{\ms}{\mathfrak{s}}

\newcommand{\hmt}{\hat{\mathfrak{t}}}
\newcommand{\hms}{\hat{\mathfrak{s}}}

\numberwithin{equation}{section}

\begin{document}

\begin{titlepage}

\rightline{\vbox{\small\hbox{\tt hep-th/0606246} }} \vskip 3cm

\centerline{\Large \bf Three-Charge Black Holes on a Circle}

\vskip 1.6cm \centerline{\bf Troels Harmark$\,^{1}$, Kristjan R.\
Kristjansson$\,^{2}$, Niels A. Obers$\,^{1}$, Peter B.
R{\o}nne$\,^{1}$} \vskip 0.5cm
\begin{center}
\sl $^1$ The Niels Bohr Institute  \\
\sl  Blegdamsvej 17, 2100 Copenhagen \O , Denmark \\
\vskip 0.5cm
\sl $^2$ Nordita \\
\sl  Blegdamsvej 17, 2100 Copenhagen \O , Denmark \\
\end{center}

\vskip 0.5cm

\centerline{\small\tt harmark@nbi.dk, kristk@nordita.dk,
obers@nbi.dk, roenne@nbi.dk}

\vskip 1.2cm \noindent
We study phases of
five-dimensional three-charge black holes with a circle in their transverse space.
In particular, when the black hole is localized on the circle we compute the corrections
to the metric and corresponding thermodynamics in the limit of small mass. When taking
the near-extremal limit, this gives the corrections to the finite entropy of
the extremal three-charge black hole
as a function of the energy above extremality. For the partial extremal limit with
two charges sent to infinity and one finite we show that the first correction to the
 entropy is in agreement with the microscopic entropy by taking into account
that the number of branes shift as a consequence of the interactions across the transverse
circle. Beyond these analytical results, we also numerically obtain the
entire phase of non- and near-extremal three- and two-charge black
holes localized on a circle. More generally, we  find in this
paper a rich phase structure, including a new phase of three-charge
black holes that are non-uniformly distributed on the circle. All these three-charge
black hole phases are found via a map that relates them to the phases of five-dimensional
neutral Kaluza-Klein black holes.


\end{titlepage}

\pagestyle{empty} \tableofcontents
\pagestyle{plain} \setcounter{page}{1}

\section{Introduction}
\label{sec:intro}

Three-charge black holes in five dimensions play a prominent role in
string theory since the Bekenstein-Hawking entropy of these black
holes can be explained by a microscopic counting of the degeneracies
of states \cite{Strominger:1996sh,Callan:1996dv,Horowitz:1996fn,Breckenridge:1996is,
Horowitz:1996ay}.
 In particular, such three-charge black holes
can be constructed by considering appropriate D-brane
configurations in ten-dimensional string theory and reducing to
five-dimensional supergravity. These three-charge black holes (as
well as the related two-charge system) continue to receive a lot
of attention in the literature, providing a fertile ground for
further exploration of the microscopic origin of entropy in string
theory (see e.g. the review \cite{Mathur:2005zp,Mathur:2005ai})
and understanding of the AdS${}_3$/CFT${}_2$ correspondence
\cite{Maldacena:1997re,Aharony:1999ti}.

One direction that we wish to explore in this paper is what
happens to the entropy of such five-dimensional three-charge black
holes when we compactify one of the directions, i.e. when we
consider the asymptotic space to be four-dimensional Minkowski
space times a circle. Obviously, for extremal black holes the BPS
property ensures that we can localize them on the circle by
considering an infinite array on the covering space of the circle.
The entropy of this extremal configuration is the well-known
constant $ 2 \pi \sqrt{N_1 N_4 N_0 }$.%
\footnote{We generally work in this paper in a duality frame in which
the extremal configuration is the F1-D0-D4 system, which is related to the P-D1-D5 system
by a T-duality in the direction of the F1-string.}
 However, when we  move away from
extremality the black holes will start to interact with each other
and as a consequence the finite entropy will receive corrections.

One of the primary aims of this paper is the computation of these
corrections in the limit of small mass (or, equivalently, large
compact radius). In particular, in the near-extremal limit we find
that the entropy including the first two corrections is given by
\begin{align}
\label{eq:bigShatrho0}  S= 2\pi \sqrt{N_1 N_4  N_0 }
    \left(1+\sqrt{\frac{\epsilon}{8}} + \frac{\epsilon}{16}
    + \mathcal{O}(\epsilon^{3/2})\right)
\end{align}
where $\epsilon$ is a dimensionless quantity proportional to the energy above
extremality. We also compute the corrected entropy when taking  a partial extremal limit
in which two of the charges are sent to infinity and one is kept finite.
In this case, we are able to show that the first correction to the  entropy is in agreement with
the microscopic entropy formula \cite{Strominger:1996sh,Horowitz:1996ay}
\begin{equation}
\label{eq:Smic}
S = 2 \pi \sqrt{N_1 N_4} \left[ \sqrt{N_0} + \sqrt{\bar{N}_0} \right]
\end{equation}
when taking into account that as a consequence of the interactions across the transverse circle
the number of D0 and anti-D0 branes will shift \cite{Costa:2000kf}.

Beyond these analytical results, we will also numerically obtain the
entire phase of non-and near-extremal three (and two-) charge black
holes localized on a circle. More generally, we will find in this
paper a rich phase structure, including a new phase of three-charge
black holes that are non-uniformly distributed on the circle.

The method we employ in this paper builds on the one used in
\cite{Harmark:2004ws} (see also
\cite{Bostock:2004mg,Aharony:2004ig}) where it was shown that any
Kaluza-Klein black hole (see \cite{Harmark:2005pp,Kol:2004ww} for
reviews) in $d+1$ dimensions ($ 4 \leq d \leq 9$) can be mapped to a
corresponding brane solution of Type IIA/IIB String Theory and
M-theory, following the procedure originally conceived in
\cite{Hassan:1992mq}. These are thermal excitations of
extremal 1/2-BPS branes in String/M-theory with
transverse space $\R^{d-1} \times S^1$, i.e.\ with a transverse
circle. This gave a precise connection between the rich phase
structure of Kaluza-Klein black holes and that of the corresponding brane.
In particular, by considering the near-extremal limit of the latter,
the thermodynamic behavior of the non-gravitational theories dual to
near-extremal branes on a circle was obtained via the map.

As we will show in detail in this paper, by considering the
particular case of Kaluza-Klein black holes in five dimensions
$(d=4)$, the above map can be generalized to generate non-extremal
three-charge brane configurations in Type IIA/IIB String Theory
and M-theory from any five-dimensional Kaluza-Klein black hole. We
will typically work in the duality frame where these charges are
carried by the F1-D0-D4 system, and the solutions obtained are
thermal excitations of the corresponding extremal 1/8-BPS brane
system. When reduced on the spatial world-volume directions of
these branes, we then obtain three-charge black holes in
five-dimensional supergravity.

While the procedure used to  generate these three-charge black holes
is a standard generalization of the one-charge case, it turns out
that many features and properties of the resulting map from
five dimensional Kaluza-Klein black holes to three-charge black holes
are very different in this case. At various points in the paper we will
comment on these differences which seem highly connected to the fact
that, contrary to the one-charge case, the extremal localized three-charge
black hole has finite entropy.

Fortunately, much is known about the phase structure of
five-dimensional Kaluza-Klein black holes. There are four known
phases: \newline \noindent i) The uniform phase, i.e. the black
string which is a four-dimensional Schwarzschild black hole times a
circle. The horizon topology in this phase is $S^2 \times S^1$.
\newline \noindent ii) The non-uniform phase, which is a static
solution emerging from the uniform phase at the Gregory-Laflamme point
\cite{Gregory:1993vy,Gregory:1994bj}
 where the black string is marginally unstable. The leading order behavior
of this phase was found in
\cite{Gubser:2001ac,Wiseman:2002zc,Sorkin:2004qq}
 and very recently the entire%
 \footnote{The non-uniform and localized phase are conjectured \cite{Kol:2002xz}
 to meet in a topology changing transition point. For Kaluza-Klein black holes
 in five and six dimensions this is highly supported by the data of
 \cite{Kudoh:2004hs,Kleihaus:2006ee}. However,
 there is still a ``gap'' between the localized and non-uniform phase,
 so that the numerical branches may not be fully complete.}
 phase was numerically computed for five dimensions in
Ref.~\cite{Kleihaus:2006ee}.
 This phase also has horizon topology $S^2 \times S^1$ but the horizon is non-uniform
 along the direction of the circle. \newline
\noindent iii) The localized phase, which approaches the five-dimensional
Schwarzschild black hole in the limit of zero mass. The first
correction to this metric in the small-mass limit is known analytically
\cite{Harmark:2003yz} (see also
\cite{Gorbonos:2004uc,Gorbonos:2005px}) using the ansatz proposed in
\cite{Harmark:2002tr}  and recently also the second correction
\cite{Karasik:2004ds,Chu:2006ce}%
\footnote{Note that the second order correction to the
thermodynamics of the localized phase was obtained in
\cite{Chu:2006ce} for all $d$. However for the case of $d=4$ in
addition the corrections to the metric are known
\cite{Karasik:2004ds}. For simplicity we use in this paper only the
first order corrected metric, but for the thermodynamics we include
these second order corrections.}, while the entire phase was
obtained numerically in \cite{Sorkin:2003ka,Kudoh:2003ki,
Kudoh:2004hs}. The horizon topology in this phase is $S^3$. \newline
\noindent iv) The phase of bubble-black hole sequences
\cite{Emparan:2001wk,Elvang:2002br,Elvang:2004iz}, involving black
objects held apart by Kaluza-Klein bubbles, where the topology of
the black object is $S^3$ or $S^2 \times S^1$ depending on the
position in the sequence. The bubbles support the contractible $S^1$
and provide the repelling force to keep the black holes in static
equilibrium. For any such alternating sequence of bubbles and black
holes the exact form of the solution is known.

The first three of the above three phases (not involving
Kaluza-Klein bubbles) can be found in the $SO(3)$ symmetric ansatz
proposed in \cite{Harmark:2002tr} and proven in
\cite{Wiseman:2002ti,Harmark:2003eg}, while the phase involving
Kaluza-Klein bubbles occur in the generalized Weyl ansatz of
Ref.~\cite{Emparan:2001wk},
 with $SO(2)^3$ symmetry. Moreover, these phases can be summarized by plotting them
 in a two-dimensional phase diagram \cite{Harmark:2003dg,Kol:2003if,Harmark:2003eg}
where (relative) tension is plotted versus the mass.

By applying a sequence of boosts and U-dualities we show in this
paper that for any five-dimensional Kaluza-Klein black hole, one obtains
a corresponding five-dimensional three-charge black hole solution
with a transverse circle. Each of the phases of Kaluza-Klein black
holes described above thus directly maps onto a corresponding phase
of non-extremal three-charge black holes and we can express the
thermodynamic properties in terms of those of the seeding solution.
Furthermore, the map becomes especially simple (as was the case for
the one-charge branes considered in Ref.~\cite{Harmark:2004ws})
when we take the near-extremal limit. This limit is also relevant
for the dual CFT description of these brane systems and for our
application to microstate counting. Various near-extremal limits are
considered, in which three, two, and one of the charges are sent to
infinity. We also discuss the special case when one of the three
charges is zero. In this case the map of the thermodynamics from the
neutral seeding solution to that of the near-extremal two-charge
solution is also very simple, in analogy to the one-charge case
considered in \cite{Harmark:2004ws} (see \cite{Harmark:2005pq} for a
very short review). In the applications discussed in this paper we
will primarily focus on the first three phases listed above. In
particular, our emphasis will be on the localized phase, but we will
also present the results for the uniform phase and new non-uniform
phase.

The localized phase of the F1-D0-D4 system is particularly
interesting, since by a T-duality in the direction of the F1-string
this is dual to the P-D1-D5 system, and hence has a finite entropy
in the extremal limit. For this case we can summarize our results as
follows:
\begin{itemize}
\item Using the analytical results of \cite{Harmark:2003yz} for the first order correction
 and the second order correction computed in \cite{Karasik:2004ds},
we find the corrected metric for non-extremal three-charge black
holes on a circle and the corresponding thermodynamics in an
expansion for small masses.
\item We consider the near-extremal limit in which all three charges are sent to infinity,
and give the corrected metric for near-extremal three-charge black
holes on a circle and the corresponding corrected thermodynamics. In
particular, we find the leading order correction \eqref{eq:bigShatrho0} to the finite
entropy of the F1-D0-D4 system due to the presence of the circle. We
also use the numerical results of \cite{Kudoh:2004hs} to plot the
thermodynamics of the entire localized phase of the F1-D0-D4 brane
system on a circle.
\item We furthermore consider special cases of the near-extremal limit in which one (or two) of the
three charges are kept finite. In particular, for the case with one
finite charge, we obtain the corrected entropy and show that the
leading order correction is in agreement with a microscopic counting
for the three-charge system on a circle in the dilute gas
approximation. This is done by using the microscopic formula \eqref{eq:Smic} and
using the fact that, for fixed total energy, the charges are shifted due to the
interaction across the transverse circle.
\item The non-extremal three-charge system includes as a special case the two-charge
D0-D4 system. By a T-duality this corresponds to the D1-D5 system
with a transverse circle, so that the localized phase in this case
is relevant for the dual two-dimensional CFT
\cite{Maldacena:1997re}. For this case we also obtain the corrected
thermodynamics for small masses, and plot the thermodynamics of the
entire localized phase. In the canonical ensemble, the corrections describe the
small temperature expansion of the free energy around the conformal behavior $F \propto T^2$
of the two-dimensional CFT.
\end{itemize}

In parallel to these applications for the localized case, which from
the point of view of microstate counting and dual CFT are most
interesting, we also discuss the charged black hole solutions that
are mapped from the uniform and non-uniform phase of
five-dimensional Kaluza-Klein black holes. The latter phase
generates a new phase of three-charge black holes non-uniformly
distributed on the circle, which is interesting in its own right as
a new static brane configuration in String/M-Theory, or equivalently
in the corresponding five-dimensional supergravity.

When all three
charges are non-zero, the uniform phase of F1-D0-D4 is T-dual to
a P-D1-D5 system uniformly smeared on the transverse circle, which in
turn can be dualized to a P-D2-D6 system. Correspondingly, the
non-uniform phase is thus equivalent to the P-D1-D5 system
non-uniformly distributed on the transverse circle. For this phase
we can use the recent numerical data \cite{Kleihaus:2006ee} for the non-uniform phase of
five-dimensional Kaluza-Klein black holes to obtain plots of the thermodynamics.
Likewise for the two-charge case, we have a new phase of D1-D5
branes non-uniformly distributed on the transverse circle. It turns
out that in the near-extremal limit the uniformly smeared phase has
Hagedorn behavior, with the non-uniform phase emerging at the
Hagedorn point. More generally, it is found that
 the thermodynamic behavior for the near-extremal two-charge case is completely analogous to that of the near-extremal
NS5-brane recently considered in Ref.~\cite{Harmark:2005dt}.

The outline of the paper is as follows. In Section
\ref{sec:3chargebh} we show how to generate ten-dimensional
three-charge configurations on a circle by applying a set of
boosts and U-duality transformations to a general five-dimensional
neutral Kaluza-Klein black hole. We will also refer to the latter
as the seeding solution. We discuss how to measure the asymptotic
quantities for these three-charge solutions and the corresponding
mapping of physical quantities from the seeding solution to those
of the non-extremal three-charge system.

In Section \ref{sec:appl} we first present some details on the
seeding solutions, restricting to those that do not involve
Kaluza-Klein bubbles, and which fall in the ansatz of
\cite{Harmark:2002tr}. As applications of the map of Section
\ref{sec:3chargebh} we then discuss the uniform, non-uniform
and localized phase of three-charge black holes.

Section \ref{sec:3chargenearextr} defines the near-extremal limit
that we take on the non-extremal three-charge solutions and
discusses the corresponding physical quantities for the resulting
near-extremal three-charge configurations when all three charges are sent to infinity.
It turns out that for these configurations, tension along the transverse
circle is proportional to the energy above extremality
(unlike the one-charge case) and we
present a general argument showing that this is the case for any
system that involves a non-vanishing entropy in the extremal limit.

We then continue in Section \ref{sec:3nearphase} by presenting the
near-extremal phases that follow by applying the near-extremal limit
of Section \ref{sec:3chargenearextr} to the phases considered in Section
\ref{sec:appl}. This includes the uniform, non-uniform and localized
phase of three-charge black holes on a circle.

Section \ref{sec:extensions} considers other
near-extremal limits, in which we keep one or two of the three
charges finite. In particular, the corrected thermodynamics for
the localized phase is computed in these two cases. Furthermore,
we discuss the special case where one of the three-charges is zero,
i.e.\ non-extremal two-charge configurations with a transverse
circle. In particular, the near-extremal limit of the D0-D4 system
on a transverse circle is considered and the general map of the
physical quantities. As an application we apply this map to the
uniform, non-uniform and localized phase and present the resulting
thermodynamics for each of these phases.

Finally, in Section \ref{sec:microstates} we turn to the specific
case of the partial extremal limit where two of the charges are
sent to infinity and one is kept finite, focussing on the localized three-charge
black hole. In this dilute gas regime we present a microstate counting argument,
following Ref.~\cite{Costa:2000kf}, that reproduces the first correction computed
in  Section \ref{sec:extensions} using the map. We end with the conclusions and outlook
in Section \ref{sec:concl}.

A number of appendices is included. Appendix \ref{sec:Uduality} gives some
useful details on the boosts and U-dualities employed in Section
 \ref{sec:3chargebh}. Appendix \ref{app:cmap} gives the relation between the asymptotics
of the seeding solution and that of the three-charge solutions.
Appendix \ref{app:elec} provides further details on our definition of
electric masses and tensions.

\section{Generating three-charge solutions from Kaluza-Klein black holes}
\label{sec:3chargebh}

In this section we will present the non-extremal three-charge
solution generated from a neutral Kaluza-Klein black hole in 4+1
dimensions.  The neutral solution is referred to as the {\em seeding
solution}.  Further, we will see how to calculate the physical
quantities of the new three-charge solution given the seeding
solution.

\subsection{Three-charge configuration on a circle}
\label{sec:3chargeconfig}

The system that we are interested in is a three-charge solution of
Type IIA Supergravity that describes a thermal excitation of the
1/8-BPS configuration with an F1-string, D4-brane and a D0-brane. The
configuration is a solution to the equation of motions of the action
\begin{equation}
\label{eq:action}
I= \frac{1}{16\pi G_{10}} \int d^{10}x\sqrt{-g}
\left( R {-} \frac{1}{2}\partial_\mu \phi \partial^\mu \phi
{-}\frac{1}{12} e^{-\phi} (dB)^2
{-}\frac{1}{2\cdot6!} e^{-\frac{1}{2}\phi} (dA_{(5)})^2
{-}\frac{1}{4} e^{\frac{3}{2}\phi} (dA_{(1)})^2
\right)
\end{equation}
where $\phi$ is the dilaton field, $B$ is the Kalb-Ramond two-form
field and the $A_{(i)}, i=5,1$ are the gauge fields that couple to
the D4-brane and the D0-brane, respectively.
This is the low energy action of Type IIA String Theory when the only
gauge fields present are the ones that correspond to the three
extended objects  that we are interested in.  Note that the action is
written in Einstein frame which we will use throughout.

The extremal 1/8-BPS solution of the action \eqref{eq:action} is
well known and can be found {\it e.g.}\ by using the harmonic rule
\cite{Horowitz:1996ay,Tseytlin1996,Gauntlett1996}. The metric for
such a solution consists of a world-volume part for the extended
objects times a transverse space which will be four-dimensional if
the extended objects do not intersect.  It is
important that the non-compact part of the transverse space has at
least three dimensions in order to be able to measure asymptotic
quantities.
We are interested in solutions where the transverse space asymptotes
to $\mathbb{R}^3\times S^1$. The compact transverse circle gives
rise to some interesting physics that we wish to explore.

The solution can be compactified on $T^5$ which is spanned by the
spatial world-volume directions of the D4-brane and the F1-string.
This gives a five-dimensional black hole with three charges that
can be compared to the extremal solution of Strominger and Vafa
\cite{Strominger:1996sh}.
We choose to consider the F1-D0-D4 configuration instead of the more
traditional P-D1-D5 configuration because the background turns out
to be simpler, it has diagonal metric and the objects do not share
spatial world-volume directions. The systems are, however, T-dual.

\subsection{Generating  F1-D4-D0 solutions}
\label{sec:adding3charges}

The main idea of the present paper is that we can generate charged
solutions of the type described above, starting from a neutral 4+1
dimensional Kaluza-Klein black hole.  A $d+1$ dimensional static
and neutral Kaluza-Klein black hole is defined here as a pure
gravity solution that has at least one event horizon and
asymptotes to $d$-dimensional Minkowski-space times a circle at
infinity.
The thermodynamics of these kinds of Kaluza-Klein black holes has
been studied extensively recently and there are both
numerical and analytical results  available about their different phases
(see the reviews \cite{Harmark:2005pp,Kol:2004ww}).
 The key observation of the present paper is that we can
translate information about the thermodynamics of the well-studied
seeding solutions into information about the thermal three-charge
configuration in ten-dimensional supergravity with a transverse circle
where little or nothing was known before.

Let us start with a static and neutral five-dimensional Kaluza-Klein
black hole as a seeding solution. There is no dilaton and no gauge
fields and we assume that the metric can be written in the form
\begin{equation}
\label{eq:metricV}
ds_5^2 = -U dt^2 + \frac{L^2}{(2\pi)^2} V_{ab}dx^a dx^b
\end{equation}
where $U$ is a non-constant function that vanishes at the horizon(s)
and asymptotes to one, and $V_{ab}dx^a dx^b$ describes a cylinder of
circumference $2\pi$ in the asymptotic region. The metric should in
other words asymptote to
\begin{align}
\label{eq:asympmetric}
ds^2 = -dt^2  + dr^2 + r^2 d\Omega_2^2 + dz^2
\end{align}
where $z$ is periodic with period $L$. We will refer to the
dimensionful coordinates $r$ and $z$ when discussing the asymptotic
behavior of the full metric.

By adding five flat dimensions $x$ and $u_i$, $i=1,...,4$, to the
neutral solution \eqref{eq:metricV}, and performing a series of
boosts and U-dualities, we can construct a ten-dimensional
solution of Type IIA Supergravity with three-charges.  Each boost
adds one charge which depends on the rapidity parameter
$\alpha$ of the boost. The derivation is sketched in Appendix
\ref{sec:Uduality}.
The new solution has metric
\begin{equation}
\label{eq:einsteinmetric}
ds_{10}^2 = H_1^{-\frac{3}{4}}H_4^{-\frac{3}{8}}H_0^{-\frac{7}{8}}
    \left( -Udt^2  +H_4H_0 dx^2 + H_1H_0 \sum_{i=1}^4 (du^i)^2 +
             H_1H_4H_0\frac{L^2}{(2\pi)^2} V_{ab}dx^a dx^b  \right),
\end{equation}
a dilaton $\phi$ given by
\begin{equation}
\label{eq:dilaton}
 e^{2\phi} = H_1^{-1}H_4^{-\frac{1}{2}}H_0^{\frac{3}{2}},
\end{equation}
a Kalb-Ramond field given by
\begin{align}\label{eq:Kalb-Ramond}
B = \coth \alpha_1 (H_1^{-1} - 1) dt \wedge dx,
\end{align}
and gauge fields
\begin{align}\label{eq:A5}
A_{(5)} &= \coth \alpha_4 (H_4^{-1} - 1) dt \wedge du^1 \wedge du^2 \wedge du^3 \wedge du^4,\\
A_{(1)} &= \coth \alpha_0 (H_0^{-1} - 1) dt.\label{eq:A1}
\end{align}
The $H_a$ are harmonic functions of the transverse space, given by
\begin{equation}
\label{eq:Ha}
H_a = 1+ (1-U) \sinh^2 \alpha_a, \quad \textrm{for }a = 1,4,0.
\end{equation}
This three-charge solution describes a non-extremal configuration
with an F1-string, D4-brane and D0-brane.  The label $a=1,4,0$ refers to the
type of object.  In the following we will refer to the fields $B$, $A_{(5)}$
and $A_{(1)}$ collectively as $A_a$.

We can perform one more U-duality and map this solution
into a configuration with three non-extremal M2-branes which is an
excitation of a known $1/8$-BPS state.

\subsection{Measuring asymptotic quantities}
\label{sec:measuring}

The physical quantities of the configuration that can be measured
asymptotically far away in the transverse space are the mass, the
three different kind of charges, and, since we have a compact circle
in the transverse space, the tension in the direction of the circle.
The tension measures how hard the black hole pulls itself across the
circle and has an interpretation as the binding energy of the black
holes in the covering space of the circle. For extremal BPS black
holes there is no net force between the black holes in the covering
space because the electric force exactly cancels the gravitational
force and the tension therefore is zero. For non-extremal black
holes this is not the case and more interesting physics appears.
We now show how
information about all these quantities can be mapped from the
neutral seeding solution to the new three-charge solution.

It is useful to assume that each spatial world-volume direction $x$ and
$u^i$ is compactified on a circle of length $L_x$ and $L_{u^i}$ respectively.
This gives us two rectangular tori with volumes $V_1$ and
$V_4 = \prod_i L_{u^i}$.
In the asymptotic region of the transverse space the components of
the metric can be expanded to leading order\footnote{Note that the
dependence on $r$ really is in terms of $r^{(d-3)}$ where $d$ is the
number of transverse dimensions which in this case is 4.}  in $r$
\begin{align}
\label{eq:barctbarcz}
g_{tt} &\simeq -1 + \frac{\bar c_t}{r} ,
    \quad g_{zz} \simeq 1 + \frac{\bar c_z}{r} \\
\label{eq:barcxbarcu}
g_{xx} &\simeq 1 + \frac{\bar c_x}{r},
    \qquad g_{ii} \simeq 1 + \frac{\bar c_u}{r}, \quad \textrm{for }i=1,...,4,
\end{align}
and the dilaton and the non-vanishing components of the gauge fields
are to leading order
\begin{align}
\label{eq:barcrbarca}
\phi \simeq \frac{\bar c_\phi}{r}, \qquad
(A_a)_{t...} \simeq \frac{\bar c_{A_a}}{r}, \quad \textrm{for } a=1,4,0.
\end{align}
The non-vanishing gauge field components are the same as in equations
\eqref{eq:Kalb-Ramond}--\eqref{eq:A1}.

The total mass and the total tension along any compact direction can be
found from the asymptotic behavior of the metric using the general formulae
of \cite{Harmark:2004ch} (see also \cite{Traschen:2001pb,Townsend:2001rg}).
Following \cite{Harmark:2004ws,Harmark:2004ch}
we find the mass and charge via
\begin{align}
\label{eq:barMbarc}
\bar M &= \frac{\Omega_2}{g L} \left( 2 \bar c_t - \bar c_z - \bar c_x - 4 \bar c_u \right) \\
\label{eq:Qabarc}
Q_a &= -\frac{\Omega_2}{g L}   \bar c_{A_a}
\end{align}
where we have defined
\begin{equation}
\label{eq:gdefined}
g \equiv \frac{16 \pi G_{10}}{V_1V_4 L^2}.
\end{equation}
The dimensionful parameter $g$ is useful to define dimensionless
mass and charge as
\begin{equation}
\label{eq:mubarqa}
\bar \mu \equiv g\bar M, \qquad q_a \equiv g  Q_a.
\end{equation}
The tensions in the compactified directions are found via \cite{Harmark:2004ch}
\begin{align}
\label{eq:LTzbarc}
L\bar {\cal T}_z
    &= \frac{\Omega_2}{g L} \left( \bar c_t - 2 \bar c_z - \bar c_x - 4 \bar c_u \right) \\
\label{eq:LxTxbarc}
L_x\bar {\cal T}_x
    &= \frac{\Omega_2}{g L} \left(  \bar c_t - \bar c_z - 2\bar c_x - 4 \bar c_u \right) \\
\label{eq:LuTubarc}
L_{u^i}\bar {\cal T}_{u^i}
    &= \frac{\Omega_2}{g L} \left(  \bar c_t - \bar c_z - \bar c_x - 5 \bar c_u \right).
\end{align}
The world-volume of the D0-brane has no spatial direction, but for calculational
purposes we can still pretend that there is a ``phantom" $u^0$ direction compactified
on a circle of length $L_0$.  The tension in that direction would then be given by
\begin{align}
\label{eq:L0T0barc}
L_0 \bar{\cal T}_0 &= \frac{\Omega_2}{g L} \left( \bar c_t - \bar
c_z - \bar c_x - 4 \bar c_u - \bar c_0 \right) .
\end{align}
This is useful when we discuss the contribution of the D0-brane to the electric mass
in the next subsection.

\subsection{Mapping of physical quantities}
\label{sec:mapping}

To find how the physical quantities of the charged solution are
related to the original seeding solution we write the asymptotics of
the metric of the seeding solution as
\begin{align}\label{eq:seedingctcz}
-g^\textrm{seed}_{tt} = U \simeq 1 - \frac{c_t}{r},  \qquad
g^\textrm{seed}_{zz} \simeq 1 + \frac{c_z}{r}.
\end{align}
Expressed in $c_t$ and $c_z$ the mass and tension of the seeding
solution is given by \cite{Harmark:2003dg,Kol:2003if}
\begin{align}
\label{eq:MTz}
    M=\frac{\Omega_{2}L}{16\pi G_5}(2c_t-c_z),\quad{\cal T}_z
       =\frac{\Omega_{2}L}{16\pi G_5}(c_t-2c_z).
\end{align}
Correspondingly we have the dimensionless mass, $\mu$, and tension,
$n$
\begin{equation}
\label{eq:mundefined}
    \mu=\frac{16\pi G_5}{L^2}M=\frac{\Omega_2}{L}(2c_t-c_z),\quad
    n=\frac{{\cal T}L}{M}=\frac{c_t-2c_z}{2c_t-c_z}.
\end{equation}

By plugging the asymptotics \eqref{eq:seedingctcz} into the solution
\eqref{eq:einsteinmetric}--\eqref{eq:Ha} and expanding to first
order, we find a relation between the expansion coefficients of the
new solution and the seeding solution.  This relation is spelled out
in Appendix \ref{app:cmap}.
Plugging into the equations for mass, charge and tension we get
\begin{align}
  \label{eq:barM}
\bar M &=  \frac{\Omega_2}{g L}
    \left( -c_z +  c_t ( 2 + \sinh^2\alpha_1
       + \sinh^2\alpha_4 + \sinh^2\alpha_0 ) \right),\\
\label{eq:Qa}
Q_a &= \frac{\Omega_2}{g L}  c_t \sinh\alpha_a \cosh\alpha_a \\
\label{eq:LTz}
L\bar {\cal T}_z &= \frac{\Omega_2}{g L}
    \left( c_t - 2 c_z \right), \\
\label{eq:LxTx}
L_x\bar {\cal T}_x &= \frac{\Omega_2}{g L}
    \left( -c_z + c_t(1 + \sinh^2 \alpha_1) \right), \\
\label{eq:LuTu}
L_{u^i}\bar {\cal T}_{u^i} &= \frac{\Omega_2}{g L}
    \left( -c_z + c_t(1 + \sinh^2 \alpha_4) \right), \\
\label{eq:L0T0}
L_0\bar {\cal T}_0 &= \frac{\Omega_2}{g L}
    \left( -c_z + c_t(1 + \sinh^2 \alpha_0) \right).
\end{align}
Thus we have found how the physical quantities of the charged solution are given
in terms of the boost parameters $\alpha_a$ and the two independent quantities
$c_t$ and $c_z$ of the neutral solution.

\subsubsection*{The electric mass and tensions}

The electric part of the mass and tensions are simply defined as the
parts that go to zero when the charges $Q_a$ vanish.
We can directly read from Equations
\eqref{eq:barM} and \eqref{eq:Qa} that
\begin{equation}
\label{eq:Mel}
M^\textrm{el}=\sum_a\frac{\Omega_2}{g L} c_t \sinh^2 \alpha_a.
\end{equation}
We see that the electric mass consists of a three parts -- one
for each of the charged objects. Thus it is natural to define the
electric mass $M^\textrm{el}_a$ corresponding to the charge
$Q_a$ as
\begin{equation}
\label{eq:Mela}
M^\textrm{el}_a = L_a  (\bar{\cal T}_a)^\textrm{el} =
\frac{\Omega_2}{g L} c_t \sinh^2 \alpha_a,
\end{equation}
for $a=1,4,0$ $(x,u^i,0)$. In this notation $a$ can either
label the type of object or one of the spatial world-volume
directions of the corresponding object.

Note that the electric mass can be written as
\begin{align}\label{eq:chemicalpotential}
 M^\textrm{el}_a = \nu_a Q_a
\end{align}
where
\begin{equation}
\label{eq:chempot} \nu_a=\tanh \alpha_a
\end{equation}
is the chemical potential. The chemical
potential can also be measured as $\nu_a=-A_{a}|_{\mathrm{Horizon}}$
which by setting $U=0$ in~\eqref{eq:Kalb-Ramond}--\eqref{eq:A1}
gives the same result.

We also note that the electric part of the tension ${\cal T}_z$ is
zero. To see how all of this follows from the harmonic function rule
in generality see Appendix~\ref{app:elec} where the electric masses
and tensions are calculated in detail.

\subsubsection*{The mapping of dimensionless quantities}

In the Kaluza-Klein black hole literature it is customary to define
a relative tension $n$ as the total tension divided by the total
mass. Branches of different types of static solutions are then
plotted on a $(\mu,n)$ phase diagram, where $\mu$ is the
dimensionless mass. This kind of phase diagrams  will be
discussed further in Section~\ref{sec:appl}.

For the charged black holes under consideration here,
we define the relative tension along the $z$ direction as \cite{Harmark:2004ws}
\begin{equation}
\label{eq:defbarn}
\bar n \equiv \frac{L \bar {\cal T}_z}{\bar M - M^\textrm{el}} =
\frac{c_t - 2c_z}{2 c_t -  c_z}
\end{equation}
and the relative tension along each of the world-volume directions as
\begin{equation}
\label{eq:defbarna}
\bar n_a \equiv \frac{L_a (\bar {\cal T}_a - \bar {\cal T}_a^\textrm{el})
                       }{\bar M - M^\textrm{el}}
    = \frac{c_t - c_z}{2c_t - c_z}.
\end{equation}
Note that we have chosen to subtract the electrical contribution in
these definitions. In Equations
\eqref{eq:defbarn}--\eqref{eq:defbarna} we have also written the
relative tensions in terms of the seeding parameters $c_t$ and
$c_z$.

\bigskip

Writing the physical parameters $\bar \mu$, $\bar n$ and $q_a$ in
terms of the original quantities $\mu$, $n$ of the seeding solution
and the boost parameters $\alpha_a$ gives
\begin{equation}
\label{eq:muintermsof}
\bar \mu = \mu \left( 1+ \frac{2-n}{3} (\sinh^2\alpha_1
    + \sinh^2\alpha_4 + \sinh^2\alpha_0) \right),
\end{equation}
\begin{equation}
\label{eq:qintermsof}
q_a = \mu \frac{2-n}{3} \sinh\alpha_a \cosh\alpha_a,
    \quad \textrm{for }a=1,4,0
\end{equation}
and finally
\begin{equation}
\label{eq:barnbarna}
\bar n = n, \qquad \bar n_a = \frac{1+n}{3} .
\end{equation}
We can solve Equation (\ref{eq:qintermsof}) for $\cosh\alpha_a$ and get
\begin{equation}
\label{eq:coshal}
\cosh\alpha_a = \sqrt{\frac{1}{2}\left(1+\frac{1}{b_a}\sqrt{1+b_a^2} \right)}
\end{equation}
with
\begin{equation}
\label{eq:ba}
b_a \equiv \frac{2-n}{6}\frac{\mu}{q_a}.
\end{equation}
Given the values of the three charges, the map between the mass and relative
tension of the neutral and charged solutions can therefore be written as
\begin{align}
\label{eq:barnisn}
\bar n &=n, \\
\bar \mu 
\label{eq:map}
& = \sum_a q_a + \frac{1}{2}\mu n
    + \frac{(2-n)\mu }{6}\sum_a \frac{b_a}{1+\sqrt{1+b_a^2}} .
\end{align}
This map from neutral to three-charge Kaluza-Klein black holes
is one of our main results.

There are a few things to notice here.  The neutral seeding solutions
always have $\mu \ge 0$ and $0\le n\le 2$  and therefore we see that
for fixed $q_a$ the mass $\bar\mu$ is bounded from below by $\sum_a
q_a$. This is to be expected for a charged black hole. An object
with mass smaller than its charge results in a naked singularity
rather than a black hole. We therefore define the {\em energy above
extremality} to be the mass minus the sum of the charges.
Later we will see that  $b_a \to 0$ in the near-extremal limit, and therefore
\begin{equation}\label{eq:energydiff}
\epsilon \equiv \bar \mu - \sum_a q_a \to \frac{1}{2}\mu n
\end{equation}
in that limit. The mass of the charged black hole can also be
written as $\bar\mu = \mu + \sum_a \nu_a q_a$, where $\nu_a =
\tanh\alpha_a$ is the chemical potential from
Equation~\eqref{eq:chemicalpotential}.

If the seeding solution has a single connected horizon we can
find its temperature $T$ and entropy $S$ from the metric.
In this paper we will mostly work with rescaled temperature and
entropy which are defined for the seeding solution as \cite{Harmark:2005pp}
\begin{align}
\label{eq:fraktfraks}
{\mathfrak t} =  LT, \quad {\mathfrak s} = \frac{16\pi G_5}{L^3} S.
\end{align}
For the three-charge solution we define the rescaled temperature
and entropy analogously by
\begin{align}
\label{eq:bartbars}
\bar {\mathfrak t} =  L \bar T, \quad
\bar {\mathfrak s} = \frac{g}{L} \bar S,
\end{align}
where $g$ is given in Equation \eqref{eq:gdefined}.
These quantities are
calculated at the horizon where $U=0$ and thus $H_a =
\cosh^2\alpha_a$. It is easy to see from the metric
(\ref{eq:einsteinmetric}) that the three-charge solution therefore
has temperature and entropy given by
\begin{align}
\label{eq:bart}
\bar {\mathfrak t} &= {\mathfrak t}/ \cosh\alpha_1 \cosh\alpha_4 \cosh\alpha_0, \\
\label{eq:bars}
\bar {\mathfrak s} &= {\mathfrak s} \cosh\alpha_1 \cosh\alpha_4 \cosh\alpha_0.
\end{align}
The factors of $\cosh\alpha_a$ cancel when we multiply the
temperature and entropy and therefore the product
\begin{equation}
\label{eq:tsts}
\bar {\mathfrak t} \bar {\mathfrak s} = {\mathfrak t} {\mathfrak s}
\end{equation}
remains fixed.
The generalized Smarr formula from \cite{Harmark:2004ch} gives a
relation between the temperature, entropy, and the gravitational
mass for our three-charge solution in terms of the relative
tension
\begin{align}
\bar {\mathfrak t} \bar {\mathfrak s} 
\label{eq:smarr}
&= \frac{2 - \bar n}{3} \left(\bar \mu-\mu^\textrm{el}\right).
\end{align}

\section{Non-extremal three-charge black holes on a circle}
\label{sec:appl}

In this section we apply the map, that was found in Section
\ref{sec:3chargebh}, to obtain three-charge black holes on a circle
from neutral Kaluza-Klein black holes. We restrict ourselves to
neutral black holes without Kaluza-Klein bubbles, and we review the
ansatz for the metric for this class of Kaluza-Klein black holes. We
then go on to describe the three different phases that we obtain for
three-charge black holes on a circle, using the map.

\subsection{The neutral seeding solutions}
\label{sec:notreview}

We have seen in Section \ref{sec:3chargebh} that we can transform
five-dimensional static and neutral black hole solutions to three-charge
solutions via boosts and U-dualities. We now specify which class of solutions
that we consider as the seeding solution, i.e.\ what solutions we
will map to three-charge solutions.

We consider the Kaluza-Klein black holes in five-dimensions (see
\cite{Harmark:2005pp,Kol:2004ww} for reviews). These are the
solutions asymptoting to four-dimensional Minkowski space times a
circle $\CM^4 \times S^1$, i.e. to Kaluza-Klein space. As reviewed
in Section \ref{sec:3chargebh} we can measure the rescaled mass
$\mu$ and relative tension $n$ for this class of solutions. In
general we have solutions with $0\leq n \leq 2$. However, the
solutions in the range $1/2 \leq n \leq 2$ have Kaluza-Klein bubbles
\cite{Elvang:2004iz,Harmark:2005pp} and we will not consider these
here. Restricting ourselves to solutions with $0 \leq n \leq 1/2$ we
have three types of solutions.
\begin{itemize}
\item The uniform black string. The metric of a uniform black
string is simply a Schwarzschild black hole times a circle. The
horizon topology is $S^2 \times S^1$.
\item The non-uniform black string. This solution is a black
string which is non-uniform around a circle. The horizon topology
is $S^2 \times S^1$. The non-uniform black string solutions have been
found numerically in \cite{Kleihaus:2006ee} (see also
\cite{Gubser:2001ac,Wiseman:2002zc,Sorkin:2004qq}).
\item Localized black holes. These are black holes which are
localized on the circle, i.e. the horizon is so small it is not
connected around the circle. The horizon topology is $S^3$.
Analytical results have been found in
\cite{Harmark:2003yz,Gorbonos:2004uc,Karasik:2004ds,Chu:2006ce}.
Numerical results have been found in \cite{Kudoh:2004hs} (see also
\cite{Sorkin:2003ka,Kudoh:2003ki}).
\end{itemize}

The metric for all three types of solutions can be written in the
ansatz \cite{Harmark:2002tr},
\begin{equation}
\label{ansatz} ds^2 = -fdt^2 + \frac{L^2}{(2\pi^2)} \left[
\frac{A}{f} dR^2 + \frac{A}{K^2} dv^2 +KR^2 d\Omega_2^2 \right] \spa
f = 1- \frac{R_0}{R} .
\end{equation}
In this ansatz the metric is specified by the two functions $A(R,v)$
and $K(R,v)$. One has furthermore that $A(R,v)$ can be found
explicitly in terms of $K(R,v)$ \cite{Harmark:2002tr}. The horizon
is located at $R=R_0$. See
\cite{Harmark:2002tr,Wiseman:2002ti,Harmark:2003eg,Harmark:2003fz}
for more on this ansatz.

\begin{figure}
\begin{center}
\includegraphics[width=0.4\columnwidth]{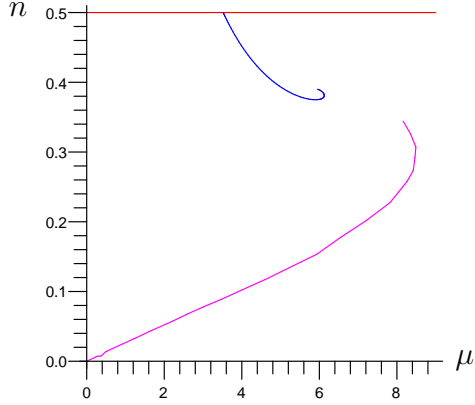}
\begin{picture}(0,0)(0,0)
\put(-6,26){\large $\mu$}\put(-175,158){\large $n$}
\end{picture}
\caption{Diagram with $\mu$ versus $n$ for the uniform black string
(red), non-uniform black string (blue) and localized black hole
phase (magenta) for five-dimensional Kaluza-Klein black holes, using
numerical results of \cite{Kudoh:2004hs,Kleihaus:2006ee}.}
\label{fig:seeding}
\end{center}
\end{figure}

We have displayed the $(\mu,n)$ phase diagram for $0 \leq n \leq
1/2$ with the three types of solutions%
\footnote{Note that for the non-uniform and localized phase there
are extra phases in the form of copies since the solutions can be
copied $k$ times on the circle
\cite{Horowitz:2002dc,Harmark:2003eg,Harmark:2004ws}.
 At the level of solutions this means that given the solution \eqref{ansatz}
  we obtain for any $k=2,3,...$
a new solution with $A'(R,v)= A(kR,kv)$, $K'(R,v) = K (k R, kv)$
and $R_0' = R_0/k$. As a consequence we have also copies for
the corresponding localized and
non-uniform phase of three-charge black holes obtained via our map.}
 in Figure \ref{fig:seeding}.
We will review these three phases further as needed for
describing the three-charge phases.

\subsection{The ansatz for three-charge black holes on a circle}
\label{sec:ansatz}

Using the map described in Section \ref{sec:adding3charges}, we map
the ansatz \eqref{ansatz} for neutral Kaluza-Klein black holes to
the following ansatz for three-charge black holes:
\begin{equation}
\label{eq:metansatz3ch}
\begin{array}{rcl}
ds_{10}^2 &=& \ds
H_1^{-\frac{3}{4}}H_4^{-\frac{3}{8}}H_0^{-\frac{7}{8}}
    \left[ -fdt^2  +H_4H_0 dx^2 + H_1H_0 \sum_{i=1}^4 (du^i)^2 \right. \\[3mm] &&
    \ds \left.
    + H_1H_4H_0\frac{L^2}{(2\pi)^2} \left(
\frac{A}{f} dR^2 + \frac{A}{K^2} dv^2 +KR^2 d\Omega_2^2 \right)
\right],
\end{array}
\end{equation}
with
\begin{equation}
\label{eq:fHansatz3ch} f = 1-\frac{R_0}{R} \spa H_a = 1+ \frac{R_0
\sinh^2 \alpha_a}{R}, \quad \textrm{for }a = 1,4,0,
\end{equation}
and with the dilaton and the gauge fields still given by
\eqref{eq:dilaton}--\eqref{eq:A1}.

Using the ansatz for three-charge black holes on a circle
\eqref{eq:metansatz3ch}, \eqref{eq:fHansatz3ch},
\eqref{eq:dilaton}--\eqref{eq:A1}, we can work out the following
explicit physical parameters (assuming a single connected horizon)
%
\begin{equation}
\label{eq:thermans}
\begin{array}{c}
\ds \bar{\mu} = 2R_0 \Big( 2-\chi+ \sum_{a} \sinh^2 \alpha_a \Big)
\spa \bar{n} = \frac{1-2\chi}{2-\chi} \spa
\\[4mm] \ds
 \bar {\mathfrak t} = \frac{1}{2 \sqrt{A_h} R_0
\prod_a \cosh \alpha_a}  \spa \bar {\mathfrak s} = 4 \sqrt{A_h}
R_0^2 \prod_a \cosh \alpha_a \spa
\\[5mm] \ds
 q_a = 2R_0 \sinh \alpha_a \cosh
\alpha_a \spa \nu_a = \tanh \alpha_a \spa \bar{n}_a =
\frac{1-\chi}{2-\chi} \spa a=1,4,0 \spa
\end{array}
\end{equation}
where $\chi$ is defined from the asymptotic behavior of $K(R,v)$ as
\begin{equation}
\label{eq:Kasy} K(R,v) = 1 - \chi \frac{R_0}{R} + \CO( R^{-2} )
\end{equation}
for $R \gg 1$, and where
\begin{equation}
\label{eq:Ahdef} A_h \equiv A(R,v) \big|_{R=R_0} \ .
\end{equation}
Note that one can show that $\partial_v A(R,v)=0$ on the horizon
$R=R_0$ \cite{Harmark:2002tr}. It is straightforward to see from the
thermodynamics \eqref{eq:thermans} along with the relation $\mu^{\rm
el} = \sum_a \nu_a q_a$ that we get the Smarr formula
\eqref{eq:smarr}.

\subsection{The uniform and non-uniform phases}
\label{sec:nuniphases}

The uniform phase corresponds to the F1-D0-D4 system smeared
uniformly on a transverse circle. The supergravity solution for this
is easily obtained by putting $A=K=1$ in the ansatz
\eqref{eq:metansatz3ch}, \eqref{eq:fHansatz3ch},
\eqref{eq:dilaton}--\eqref{eq:A1}. The thermodynamics is obtained
from \eqref{eq:thermans} setting $\chi=0$ and $A_h=1$, giving
\begin{equation}
\label{eq:thermouni} \bar{\mu} = 2R_0 \Big( 2 + \sum_{a} \sinh^2
\alpha_a \Big)  \spa
 \bar {\mathfrak t} = \frac{1}{2 R_0
\prod_a \cosh \alpha_a}  \spa \bar {\mathfrak s} = 4 R_0^2 \prod_a
\cosh \alpha_a
\end{equation}
with $q_a$ and $\nu_a$  as given in \eqref{eq:thermans}. We have
furthermore that the relative tension is $\bar{n} = 1/2$. The
uniform phase is mapped from the neutral uniform black string in
five dimensions. The horizon topology for our non-extremal F1-D0-D4
system in the uniform phase is $T^5 \times S^2 \times S^1$, where
the $T^5$ is along the charged directions.

The non-uniform phase of our F1-D0-D4 system is a phase in which the
F1-D0-D4 is still distributed on the transverse circle without gaps,
but with the distribution being non-uniform along the circle
direction. The horizon topology is therefore the same as for the
uniform phase: $T^5 \times S^2 \times S^1$. The non-uniform phase is
mapped by \eqref{eq:einsteinmetric}--\eqref{eq:Ha} from the neutral
non-uniform black string phase. From this fact it is easy to see
using Eq.~\eqref{eq:map} that the non-uniform phase emanates from
the uniform phase in a critical point corresponding to the mass
\begin{equation}
\label{eq:barmuc} \bar{\mu}_c = \sum_a q_a + x + x^2 \sum_a (q_a +
\sqrt{x^2+q_a^2})^{-1} \spa x \simeq 0.88
\end{equation}
which is mapped from the Gregory-Laflamme mass $\mu_{\rm GL} = 3.52$
of the five-dimensional neutral uniform black string
\cite{Gubser:2001ac,Sorkin:2004qq,Harmark:2005pp} using that
$x=\mu_{\rm GL} /4$. We expect that the uniform phase is unstable to
linear perturbations for masses $\bar{\mu} < \bar{\mu}_c$. Indeed,
one should be able to find the explicit unstable mode using the
methods of \cite{Aharony:2004ig,Harmark:2005jk} where
the unstable mode of one-charge smeared branes were constructed
by transforming the unstable mode for the neutral black string.

As reviewed in Section \ref{sec:notreview}, the neutral non-uniform
black string solution is obtained numerically in
\cite{Kleihaus:2006ee}. This numerical solution can then be mapped
to a numerical solution for the non-uniform phase of the F1-D0-D4
system, using either the map
\eqref{eq:einsteinmetric}--\eqref{eq:Ha}, or the ansatz
\eqref{eq:metansatz3ch}, \eqref{eq:fHansatz3ch},
\eqref{eq:dilaton}--\eqref{eq:A1}. Similarly we can map the physical
quantities using the results of Section \ref{sec:mapping}. We do not
go into details with this, since the qualitative features of the
mapped solution are highly similar to that of the neutral seeding
solution. Only in the near-extremal limits that we consider in
Sections \ref{sec:3chargenearextr} and \ref{sec:3nearphase}, one
sees significant differences in the qualitative behavior. However,
it is interesting to find the slope of the non-uniform phase in the
$(\bar{\mu},\bar{n})$ diagram near the critical point $(\bar{\mu}_c,1/2)$
since this in a simple way can tell us about some of the features of
the non-uniform phase as we change the charges. Using that the
neutral non-uniform black string has the slope $n \simeq 1/2 - \gamma
(\mu-\mu_{\rm GL})$ ($\gamma \simeq 0.14$) near the Gregory-Laflamme point
$(\mu,n)=(\mu_{\rm GL},1/2)$
\cite{Gubser:2001ac,Sorkin:2004qq,Harmark:2005pp}, we get the slope
\begin{equation}
\label{eq:nonslope} \bar{n} \simeq \frac{1}{2} - \eta
(\bar{\mu}-\bar{\mu}_c) \spa
\eta = \gamma \left[ \frac{1}{4} - 2 \gamma x + x \left( \frac{1}{4} + \frac{2}{3} \gamma x \right)
\sum_a  \frac{ 2 q_a \sqrt{x^2+q_a^2} + x^2 + 2 q_a^2}{\sqrt{x^2+q_a^2} (
q_a + \sqrt{x^2 + q_a^2})}  \right]^{-1}
\end{equation}
for $0 \leq \bar{\mu}-\bar{\mu}_c \ll 1$, when considering fixed
charges $q_a$.

\subsection{The localized phase}
\label{sec:locphase}

\newcommand{\rht}{\tilde{\rho}}
\newcommand{\tht}{\tilde{\theta}}

The localized phase of the F1-D0-D4 system corresponds to having the
horizon of F1-D0-D4 localized on the transverse circle, such that
the horizon is not connected across the circle. The horizon topology
is therefore $S^3 \times T^5$, where $T^5$ is along the charged
directions.

If we consider the case in which the size of the horizon is very
small compared to the circumference of the transverse circle, we can
write down an analytic expression for the metric using
\cite{Harmark:2003yz} (see
\cite{Gorbonos:2004uc,Karasik:2004ds,Chu:2006ce} for more analytical
results for such black holes). To this end, we should slightly
modify the ansatz \eqref{eq:metansatz3ch}, \eqref{eq:fHansatz3ch},
\eqref{eq:dilaton}--\eqref{eq:A1} by expressing it instead in the new
coordinates $\rht$ and $\tht$ defined by
\begin{equation}
\label{coordnew} 2R = \rht^2 \spa v = \pi - 2\tht + 2 \sin \tht \cos
\tht \ .
\end{equation}
We thus have the relation $\rho_0^2 = 2 R_0$ for the horizon radius.
With this, we can rewrite the ansatz \eqref{eq:metansatz3ch},
\eqref{eq:fHansatz3ch}, \eqref{eq:dilaton}--\eqref{eq:A1} as
\begin{equation}
\label{eq:metansatz3chtilde}
\begin{array}{rcl}
ds_{10}^2 &=& \ds
H_1^{-\frac{3}{4}}H_4^{-\frac{3}{8}}H_0^{-\frac{7}{8}}
    \left[ -fdt^2  +H_4H_0 dx^2 + H_1H_0 \sum_{i=1}^4 (du^i)^2 \right. \\[3mm] &&
    \ds \left.
    + H_1H_4H_0\frac{L^2}{(2\pi)^2} \left(
\frac{\tilde{A}}{f} d\rht^2 + \frac{\tilde{A}}{\tilde{K}^2} \rht^2
d\tht^2 +\tilde{K}\rht^2 \sin^2 \tht d\Omega_2^2 \right) \right],
\end{array}
\end{equation}
with
\begin{equation}
\label{eq:fHansatz3chtilde} f = 1-\frac{\rho_0^2}{\rht^2} \spa H_a =
1+ \frac{\rho_0^2 \sinh^2 \alpha_a}{\rht^2}, \quad \textrm{for }a =
1,4,0.
\end{equation}
Using the results of \cite{Harmark:2003yz} one can now write down
the full solution for the case in which the horizon is very small,
i.e. for $\rho_0 \ll 1$. For simplicity, we discuss here only the
part concerning the solution near the horizon, but it is
straightforward to use the map to find the full solution. From
\cite{Harmark:2003yz} we get for $\rho_0 \leq \rht \ll 1$ that
\begin{equation}
\label{eq:locsol} \tilde{A}^{-\frac{1}{3}} = \tilde{K}^{-1} =
\frac{1-w^2}{w} \frac{\rht^2}{\rho_0^2} + w \spa w = 1 +
\frac{1}{24} \rho_0^2 + \CO( \rho_0^4 ).
\end{equation}
{}From \cite{Harmark:2003yz} we have furthermore that $\chi =
\frac{1}{2} - \frac{1}{32} \rho_0^2+ \CO( \rho_0^4 )$ and
$\tilde{A}_h = 1 + \frac{1}{8}\rho_0^2+ \CO( \rho_0^4 )$, with
$\tilde{A}_h = \tilde{A}|_{\rht=\rho_0}$. Using then that
$A=\tilde{A}/\rht^2$ together with the thermodynamics
\eqref{eq:thermans}, we can get the thermodynamics for $\rho_0 \ll
1$. However, before writing down this thermodynamics, we note that the
second order correction has been obtained in
\cite{Karasik:2004ds,Chu:2006ce} which we can
translate to our notation as%
\footnote{Note that this follows from the slope of $n(\mu) = \mu/6^2 -
\mu^2/6^4 + \CO(\mu^3)$ for the neutral seeding solution, as one can
see from the fact that $\mu = (2-\chi) \rho_0^2$ and
$n=(1-2\chi)/(2-\chi)$.}
\begin{equation}
\label{eq:chisqrtA} \chi = \frac{1}{2} - \frac{1}{32} \rho_0^2 +
\CO(\rho_0^6) \spa \sqrt{\tilde{A}_h} = 1 + \frac{1}{16} \rho_0^2 +
\frac{1}{512} \rho_0^4 + \CO(\rho_0^6).
\end{equation}
This now gives the thermodynamics
\begin{equation}
\label{eq:loctherm}
\begin{array}{c} \ds
\bar \mu = \rho_0^2 \left( \frac{3}{2} + \frac{1}{32} \rho_0^2 +
\CO(\rho_0^6) + \sum_a \sinh^2 \alpha_a \right) \spa \bar n =
\frac{1}{24} \rho_0^2 - \frac{1}{1152} \rho_0^4 + \CO(\rho_0^6) \spa
\\[6mm] \ds
 \bar {\mathfrak t} = \frac{1- \frac{1}{16}
\rho_0^2 + \frac{1}{512} \rho_0^4 + \CO(\rho_0^6)}{\rho_0 \prod_a
\cosh \alpha_a }\spa \bar {\mathfrak s} = \rho_0^3 \left( 1 +
\frac{1}{16} \rho_0^2 + \frac{1}{512} \rho_0^4 + \CO(\rho_0^6)
\right) \prod_a \cosh \alpha_a \spa
\\[4mm] \ds
 q_a = \rho_0^2 \cosh \alpha_a \sinh \alpha_a
\spa \nu_a = \tanh \alpha_a \spa \bar n_a = \frac{1}{3} +
\frac{1}{72}\rho_0^2 - \frac{1}{3456} \rho_0^4 + \CO(\rho_0^6).
\end{array}
\end{equation}
This is thus the thermodynamics of a small three-charge black hole
localized on a circle.


\section{Near-extremal three-charge black holes on a circle}
\label{sec:3chargenearextr}

We now  consider the
near-extremal limit of our three-charge black holes on a circle. This is an interesting
limit in view of the microscopic counting of entropy and, more generally, in the context
of the dual CFT. In this section we will see how to define this limit and its
consequences for the physical quantities.

\subsection{The near-extremal limit}
\label{sec:nearextremallimit}

There are two ways to take the near-extremal limit.  One can
either keep the charges fixed and send the temperature to
zero or keep the temperature fixed and send the charges to
infinity.  Since we are interested in the thermodynamics of the
near-extremal black hole, it is natural for us to take the second option.

In order to retain the non-trivial physics related to the presence of the circle we
want to take the limit in such a way that the size of the circle has the same scale
as the energy above extremality.
This means that the metric components multiplying
$dt^2$ and $V_{ab}dx^ad x^b$ should scale in the same way.
{} From the metric in Eq.\ (\ref{eq:einsteinmetric}) we therefore require
\begin{equation}
\label{eq:nearextremalcondition}
\lim_{L\to0} H_1H_4H_0 \left(\frac{L}{2\pi}\right)^2 = \textrm{finite}.
\end{equation}
A natural way to achieve~\eqref{eq:nearextremalcondition} is to
demand
\begin{equation}\label{eq:nearextremalconditiondetailed}
    \lim_{L\to0} H_a \left(\frac{L}{2\pi}\right)^{2\gamma_a} =
\textrm{finite},\quad a=1,4,0,
\end{equation}
where $\gamma_a\geq0$ and  $\gamma_1+\gamma_4+\gamma_0=1$.
In this section we consider
only the case when all $\gamma_a$ are non-vanishing, postponing other limits
to the next section.  For $\gamma_a>0$ the requirement of Equation
\eqref{eq:nearextremalconditiondetailed} means that
\begin{equation}
\label{eq:Hadiverge}
H_a = 1+(1-U)\sinh^2\alpha_a \to \infty
\end{equation}
so that the boost parameters $\alpha_a$ must go to infinity  (for
simplicity we will take $\al_a$ to be positive). In order to see
what this means for the charges, it is convenient to introduce
rescaled coordinates on the transverse space\footnote{Note that the
coordinates of the ansatz \eqref{ansatz} approach these
dimensionless coordinates in the asymptotic region, $R\to \hat r$
and $v\to \hat z$.}
\begin{align}
\label{eq:hatrhatz}
\hat r  \equiv  \frac{2\pi}{L} r, \qquad
\hat z  \equiv \frac{2\pi}{L} z,
\end{align}
and corresponding rescaled expansion coefficients for the seeding
metric~\eqref{eq:seedingctcz}
\begin{align}
\label{eq:hatcthatcz}
-g_{tt}^\textrm{seed}  \simeq 1 - \frac{\hat c_t}{\hat r}, \qquad
g_{zz}^\textrm{seed} \simeq 1 + \frac{\hat c_z}{\hat r}.
\end{align}
These coefficients are more appropriate in the near-extremal limit since they
remain finite.
We can now write the dimensionless charges \eqref{eq:mubarqa} as
\begin{equation}
\label{eq:qa}
q_a 
= \frac{\Omega_2}{2\pi} \hat c_t \sinh\alpha_a \cosh\alpha_a
\end{equation}
and from this expression it is apparent how the charges diverge.
Notice that in order to satisfy the condition \eqref{eq:nearextremalcondition},
we should keep fixed the parameters
\begin{equation}\label{eq:ella}
    \ell_a \equiv L^{\gamma_a} \sqrt{q_a}
    = \left(\frac{\Omega_2 \hat c_t}{2\pi}\right)^{1/2}
    L^{\gamma_a}\sqrt{\sinh\alpha_a \cosh\alpha_a}.
\end{equation}

To get a finite solution for the metric, the gauge fields and the
dilaton in the near-extremal limit, we must rescale the fields
with appropriate powers of $L/2\pi$ and the powers will depend
on $\gamma_a$. This rescaling should be a symmetry of the action
\eqref{eq:action}.
It is fairly easy to check that the following scaling works
\begin{align}
\label{eq:phinew}
e^{2\phi^\textrm{new}}
    &=\left(\frac{L}{2\pi}\right)^{-2\gamma_1-\gamma_4+3\gamma_0}
    e^{2\phi^\textrm{old}}, \\
\label{eq:gttnew}
g_{tt}^\textrm{new} &=
\left(\frac{L}{2\pi}\right)^{-\frac{3}{2}\gamma_1-\frac{3}{4}\gamma_4-\frac{7}{4}\gamma_0}
g_{tt}^\textrm{old},
\qquad
g_{xx}^\textrm{new} =
\left(\frac{L}{2\pi}\right)^{-\frac{3}{2}\gamma_1-\frac{3}{2}\gamma_4+\frac{1}{4}\gamma_0}
g_{xx}^\textrm{old}, \\
\label{eq:guunew}
g_{u_iu_i}^\textrm{new} &=
\left(\frac{L}{2\pi}\right)^{\frac{1}{2}\gamma_1-\frac{3}{4}\gamma_4+\frac{1}{4}\gamma_0}
g_{u_iu_i}^\textrm{old},
\qquad
g_{rr}^\textrm{new} =
\left(\frac{L}{2\pi}\right)^{\frac{1}{2}\gamma_1+\frac{5}{4}\gamma_4+\frac{1}{4}\gamma_0+2}
g_{rr}^\textrm{old}, \\
\label{eq:Anew}
A_a^\textrm{new} &= \left(\frac{L}{2\pi}\right)^{-2\gamma_a} A_a^\textrm{old},
\qquad \qquad \quad
G_{10}^\textrm{new} = \left(\frac{L}{2\pi}\right)^{2} G_{10}^\textrm{old}.
\end{align}
The components of the metric for the other transverse directions are
rescaled in the same way as $g_{rr}$.
The choice of powers of $L/2\pi$ is unique and can be found by
first requiring the gauge fields, $B$ field, and dilaton to be
finite in the limit. Next the scalings of the metric is found by
requiring all the terms in the action to scale in the same way.
Finally, the scaling of Newton's constant is found by requiring
the scaling to be a symmetry of the action.

The choice of gauge in Equations
\eqref{eq:Kalb-Ramond}--\eqref{eq:A1} is not convenient in the
near-extremal limit because the constant term will be dominant.  But we
are free to change the gauge by adding the constant $\coth\alpha_a$
to our old $A_a^\textrm{old}$ before taking the limit. This gives
\begin{equation}
\label{eq:Aold}
A_a^\textrm{old} = \coth\alpha_a H_a^{-1} =
\left( \frac{L}{2\pi}\right)^{2\gamma_a} \coth\alpha_a \hat H_a^{-1}
\end{equation}
where we have defined
\begin{equation}
\label{eq:hatH}
\hat H_a \equiv \lim_{L \to 0} \left( \frac{L}{2\pi}\right)^{2\gamma_a} H_a .
\end{equation}
By construction, $\hat H_a$ is finite in the near-extremal limit
and from Equation \eqref{eq:Anew} we see that after rescaling
$A_a^\textrm{new}$ will be finite as well.

To summarize, the particular near-extremal limit that we are
interested in can be defined as
\begin{align}
\label{eq:nearextremallimit}
L \to 0, \quad
\alpha_a \to \infty, \quad
\ell_a \equiv L^{\gamma_a} \sqrt{q_a} = \textrm{fixed}, \quad
g \equiv \frac{16 \pi G_{10}}{V_1V_4 L^2} = \textrm{fixed}.
\end{align}

The near-extremal limit of the three-charge solution
\eqref{eq:einsteinmetric}--\eqref{eq:A1} can now be written down.
The metric and the dilaton are given by
\begin{align}
\label{eq:nearextremalmetric}
ds^2 &= \hat H_1^{-\frac{3}{4}}\hat H_4^{-\frac{3}{8}}\hat H_0^{-\frac{7}{8}}
    \left( -Udt^2  +\hat H_4\hat H_0 dx^2 + \hat H_1\hat H_0 \sum_{i=1}^4 (du^i)^2 +
             \hat H_1\hat H_4\hat H_0  V_{ab}dx^a dx^b  \right),
\end{align}
\begin{equation}
\label{eq:nearextremaldilaton}
e^{2\phi} = \hat H_1^{-1}\hat H_4^{-\frac{1}{2}}\hat
H_0^{\frac{3}{2}},
\end{equation}
where from \eqref{eq:hatH}
\begin{equation}
\label{eq:Hafork}
\hat H_a =\left\{ \begin{array}{ll} \hat h_a \frac{1-U}{\hat c_t}
& \textrm{for }\gamma_a>0\\H_a & \textrm{for }\gamma_a=0
\end{array}\right.
, \qquad
\hat h_a \equiv \frac{(2\pi)^{1-2\gamma_a} \ell_a^{2}}{\Omega_2}.
\end{equation}
In the case that all $\ga_a>0$ the dilaton is constant
\begin{equation}
\label{eq:dilconstant}
e^{2\phi} = \hat h_1^{-1}\hat h_4^{-\frac{1}{2}}\hat
h_0^{\frac{3}{2}}.
\end{equation}
The non-vanishing components of the gauge fields are given by
\begin{equation}
\label{eq:Aafork}
(A_a)_{t...} =\left\{ \begin{array}{ll} \hat H_a ^{-1} & \textrm{for
}\gamma_a>0,\\\coth \alpha_a (H_a^{-1} - 1) & \textrm{for
}\gamma_a=0.
\end{array}\right.
\end{equation}

Note that the near-horizon limit of the extremal three-charge metric is
 $AdS_2\times S^3\times T^5$ for the localized phase of F1-D0-D4. For the
uniformly smeared phase there is not such a simple description.

\subsubsection*{Relation to string scale units}
\label{sec:stringscale}

Before discussing the physical quantities of the new near-extremal solution,
it is useful to see how the parameters that are kept fixed in near-extremal
limit, namely $g$ and $\ell_a$, are related to the string coupling $g_s$ and
the string length $\ell_s$.  By comparing the parameters $\hat h_a$ in Equation
\eqref{eq:Hafork} to the usual harmonic functions of smeared extremal branes,
we find
\begin{align}
\label{eq:ell12}
\ell^2_1 &= L^{2\gamma_1} \frac{(2\pi \ell_s)^6 g_s^2 N_1}{L^2V_4} \spa \\
\label{eq:ell42}
\ell^2_4 &= L^{2\gamma_4} \frac{(2\pi \ell_s)^3 g_s N_4}{L^2V_1} \spa \\
\label{eq:ell02}
\ell^2_0 &= L^{2\gamma_0} \frac{(2\pi \ell_s)^7 g_s N_0}{L^2V_1 V_4} \spa
\end{align}
and
\begin{align}
\label{eq:ggstring}
g = \frac{(2\pi)^7\ell_s^8 g_s^2}{L^2V_1V_4}
\end{align}
where $N_1$ is the number of $F1$-strings, $N_4$ is the number of
D4-branes and $N_0$ is the number of D0-branes.

{} From the fixed parameters in Equations \eqref{eq:ell12}--\eqref{eq:ggstring}
we can form the dimensionless combination
\begin{align}
\label{eq:lllg}
\frac{\ell_1\ell_4\ell_0}{g} = 2\pi \sqrt{N_1N_4N_0}
\end{align}
which will be useful to reinstate the units in the rescaled entropy
that we obtain below.

\subsection{Physical quantities}
\label{sec:physicalquant}

In the near-extremal limit we define the energy above extremality
and the tensions in the compact directions as\footnote{We will
always assume the $Q_a$s to be positive.}
\begin{align}\label{eq:nearextremalET}
E = \lim_{L\to0} \left(\bar M - \sum_a Q_a\right), \quad
\hat{\mathcal T}_z = \lim_{L\to0} \frac{L}{2\pi}\bar {\mathcal T}_z,
\quad L_a \hat{\mathcal T}_a = \lim_{L\to0} \left( L_a\bar {\mathcal
T}_a -Q_a\right).
\end{align}
The general definitions for the energy and tensions in backgrounds
that are not asymptotically flat can be found in
\cite{Harmark:2004ch} (see also \cite{Myers:1999ps}).
In reference \cite{Harmark:2004ws} it was
actually shown that for one-charge solutions written in the ansatz
\eqref{ansatz} the general definition is equivalent
to~\eqref{eq:nearextremalET}. We have checked that the same is true
for the three-charge case.

The dimensionless versions of these quantities are defined as
\begin{align}
\label{eq:neardimless}
\epsilon = gE,
\quad \qquad r = \frac{2\pi \hat {\mathcal T_z}}{E},
\quad \qquad r_a = \frac{L_a \hat {\mathcal T}_a}{E}.
\end{align}
These variables are the possible independent physical
parameters analogous to $\mu$ and $n$ in the neutral case and $\bar
\mu$, $\bar n$ and $\bar n_a$ in the non-extremal case.

Using Equations \eqref{eq:barM} and \eqref{eq:Qa} and the fact that
\begin{equation}
\label{eq:limsinh}
\lim_{\alpha\to\infty}\left( \sinh^2\alpha - \sinh\alpha\cosh\alpha\right) = -\frac{1}{2}
\end{equation}
we find the energy above extremality in terms of $\hat c_t$ and $\hat c_z$ as
\begin{align}
\label{eq:E}
E &=  \frac{\Omega_2}{2\pi g} \left(\frac{1}{2} \hat c_t -\hat c_z\right)
\end{align}
while the tension of the transverse circle and the tensions in the
spatial world-volume directions are given by
\begin{equation}
\label{eq:2piTz}
2\pi \hat {\mathcal T}_z = \frac{\Omega_2}{2\pi g}(\hat c_t - 2\hat c_z),
\quad
L_a \hat  {\mathcal T}_a
= \frac{\Omega_2}{2\pi g}\left(\frac{1}{2}\hat c_t - \hat c_z\right).
\end{equation}

Rewriting the dimensionless energy $\epsilon$ and the
relative tensions, $r$ and $r_a$, in terms of the seeding $\mu$ and $n$
we get the remarkable result
\begin{equation}
\label{eq:erra}
\epsilon = \frac{1}{2} \mu n, \qquad r = 2, \qquad r_a = 1 \spa a =1,4,0 \ .
\end{equation}
Note that the relative tensions are constant. That means that the
tensions are proportional to the energy above extremality.  This is
a very special result that depends on the fact that we have exactly
three charges and four spatial transverse dimensions.  In Section
\ref{sec:finiteentropy}  we will see that $r$ being a constant is a
necessary condition in order for the localized five-dimensional
black hole to have a finite non-vanishing entropy in the extremal
limit. But first we look at the near-extremal thermodynamics.

The near-extremal temperature and entropy are given by
\begin{align}
\label{eq:bigThatbigShat}
\hat T = \lim_{L\to0} \bar T, \qquad \hat S = \lim_{L\to0} \bar S
\end{align}
where $\bar{T}$ and $\bar{S}$ are the non-extremal temperature and entropy.
To get rescaled temperature and entropy we need a new length
scale and it turns out to be useful to define
\begin{align}
\label{eq:ell}
\ell \equiv \ell_1 \ell_4 \ell_0
\end{align}
with $\ell_a$ given in Equation \eqref{eq:ella}.
Note that $\ell$ has the dimension of length since the $\gamma_a$
sum to one.  Dimensionless versions of the temperature and entropy
in the near-extremal limit can now be defined  by\footnote{This is
assuming that all the charges are non-zero. If one or two of
the charges are zero then the corresponding $\ell_a$ should be
left out of the definition of $\ell$. We will come back to this
in Section \ref{sec:extensions}.}
\begin{equation}
\label{eq:smallthatsmallshat}
\hat {\mathfrak t} = \ell \hat T, \qquad \hat {\mathfrak s} =\frac{g}{\ell} \hat S .
\end{equation}
These quantities can be related to the non-extremal temperature and entropy
via
\begin{align}
\label{eq:tbarqthat}
\bar {\mathfrak t} \sqrt{q_1q_4q_0} &= L \bar T \sqrt{q_1q_4q_0}
 \to \ell \hat T =  \hat {\mathfrak t} ,\\
\label{eq:}
\bar {\mathfrak s}/\sqrt{q_1q_4q_0} &= \frac{g}{L\sqrt{q_1q_4q_0}} \bar S
 \to \frac{g}{\ell}\hat S = \hat {\mathfrak s},
\end{align}
and this implies the map
\begin{equation}
\label{eq:tststs}
\hat {\mathfrak t} \hat {\mathfrak s}
    = \bar {\mathfrak t} \bar {\mathfrak s} = {\mathfrak t}{\mathfrak s}.
\end{equation}
Given the temperature and entropy of the neutral seeding solution,
we find the rescaled temperature and entropy of the near-extremal
three-charge solution as
\begin{equation}
\label{eq:thatintermsof} \hat {\mathfrak t} = {\mathfrak t}
({\mathfrak t}{\mathfrak s})^{3/2},\quad \hat {\mathfrak s} =
{\mathfrak s} ({\mathfrak t}{\mathfrak s})^{-3/2}
\end{equation}
This can be derived from
\begin{equation}
\label{eq:hattbartq}
\hat {\mathfrak t} = \bar {\mathfrak t} \sqrt{q_1q_4q_0}  =
\frac{\sqrt{q_1q_4q_0}}{\cosh\alpha_1\cosh\alpha_4\cosh\alpha_0} {\mathfrak t}
\end{equation}
by noticing that from the neutral Smarr formula $ \mt \ms = (2-n) \mu/3$ and Equation
(\ref{eq:qintermsof}) we have
\begin{align}
\label{eq:limqcosh}
\lim_{\alpha_a \to \infty} \frac{\sqrt{q_a}}{\cosh\alpha_a}
= \sqrt{{\mathfrak t} {\mathfrak s}}.
\end{align}

In the near-extremal three-charge case we do not have a
Smarr relation in the traditional sense since the relative tensions
are constant. However, we can write a Smarr relation in a `mixed'
notation where we use the relative tension $n$ of the seeding solution
\begin{align}
\label{eq:NESmarr} \hat {\mathfrak t}\hat {\mathfrak s} =
\frac{2(2-n)}{3n}\epsilon .
\end{align}
{}From this and the first law of thermodynamics we obtain
\begin{align}
\label{eq:dlogsdloge} \frac{\delta \log {\hms}}{\delta \log \hat
{\epsilon}}
    = \frac{3n}{2(2-n)}
\end{align}
so that given the curve $n(\epsilon)$ we can find the entire thermodynamics.

The Helmholtz free energy is
\begin{align}
\label{eq:freeenergy}
\hat {\mathfrak f} = \epsilon - \hat{\mathfrak t} \hat {\mathfrak s}, \quad
\delta \hat {\mathfrak f} = - \hat {\mathfrak s}\delta \hat {\mathfrak t}
\end{align}
and using the Smarr relation \eqref{eq:NESmarr} we can rewrite this
for near-extremal black holes on a circle as
\begin{align}
\label{eq:freemixed}
\hat {\mathfrak f} = \frac{5n-4}{3n}\epsilon.
\end{align}
This is the near-extremal free energy written in mixed notation, using
the neutral tension $n$ instead of $r$ which is a constant in this case.

Note that the free energy is negative for $n\le4/5$.
This is important for the dual field theory which is only
thermodynamically stable if the free energy is negative.  The region
$n\le 4/5$ contains all the usual phases with $SO(3)$
symmetry (which have $n \le 1/2$), and also
some of the Kaluza-Klein bubbles \cite{Harmark:2005pp}.

{}From the first law of thermodynamics we get
\begin{align}
\label{eq:dlogfdlogt}
\frac{\delta \log \hat {\mathfrak f}}{\delta \log \hat {\mathfrak t}}
    = -\frac{\hat {\mathfrak s}\hat {\mathfrak t}}{\hat {\mathfrak f}}
    = \frac{4-2n}{4-5n}.
\end{align}
Given $n(\hat {\mathfrak t})$ we can integrate the above equation
and get $\hat {\mathfrak f}$ as a function of $\hat {\mathfrak t}$.
Note that we again have to use the relative tension of the seeding
solution.

The world-volume pressure is
\begin{align}
\label{eq:hatpa}
\hat {\mathfrak p}_a = -r_a \epsilon = -\epsilon
\end{align}
where we used \eqref{eq:erra}. This not proportional to the free energy
\eqref{eq:freemixed},
contrary to the one-charge solutions  \cite{Harmark:2004ws} for which
the world-volume pressure is always equal to minus the free energy.

\subsection{Finite entropy from the first law of thermodynamics}
\label{sec:finiteentropy}

In this section we try to understand why the relative tension is a constant for
the near-extremal three-charge solution.  Let us start with an ansatz for the
energy above extremality in terms of the seeding $\mu$ and $n$
\begin{align}\label{eq:ansatzepsilon}
\epsilon = (a+bn)\mu
\end{align}
where $a$ and $b$ depend on the number of charges and the
number of transverse dimensions.
We can argue for this ansatz using only the expression for the
gauge fields~\eqref{eq:Kalb-Ramond}--\eqref{eq:A1}. Since $\mu$ and
$\mu n$ are linear combinations of the seeding $c_t$ and $c_z$
it is enough to show that the same is true for the energy above extremality.
To see that, we write
\begin{equation}
\label{eq:epsmunuq}
    \epsilon=\mu+\sum_a(\nu_a-1)q_a
\end{equation}
where the chemical potential $\nu_a=-A_{a}|_{\mathrm{Horizon}}$
is independent of $c_t$ and $c_z$. Since $q_a$ is
proportional to $c_t$ we immediately see that $\eps$ is indeed a
linear combination of $c_t$ and $c_z$.

In the next section we will see that $\epsilon$ takes the form
\eqref{eq:ansatzepsilon} for the one- and two-charge cases with non-zero $a$ and $b$
[cf.\ Eq.~\eqref{eq:map1} and \eqref{eq:map2} for the near-extremal map
in the one- and two-charge case respectively]
but for the three-charge case we have seen that $a=0$ [cf.\ (\ref{eq:erra})].
We will ignore this knowledge for now, and first examine what the
first law of thermodynamics implies.

{}From the non-extremal Smarr formula \eqref{eq:smarr} and the map
\eqref{eq:tststs} we know that the product of the rescaled entropy
and temperature is given by
\begin{align}
\label{eq:smarransatz}
\hat {\mathfrak t} \hat {\mathfrak s}
    =  \frac{2-n}{3} \mu
        =  \frac{2-n}{3(a+bn)}\epsilon
\end{align}
where in the last equation we used the ansatz \eqref{eq:ansatzepsilon}.
Plugging this into the first law of thermodynamics,
$\delta\epsilon = \hat {\mathfrak t} \delta \hat {\mathfrak s}$,
we therefore find
\begin{align}
\label{eq:logsloge}
\frac{\delta \log \hat {\mathfrak s}}{\delta \log \epsilon}
    = \frac{3(a+bn)}{2-n}.
\end{align}
For small black holes in the localized phase, $n\to 0$ as $\epsilon\to0$,
since the tension should vanish in the extremal (BPS) limit.
Therefore it follows that for small black holes close to extremality
\begin{align}
\label{eq:dlogssmallbh}
\frac{\delta \log \hat {\mathfrak s}}{\delta \log \epsilon} \simeq \frac{3a}{2}.
\end{align}
Integrating this equation for small $\epsilon$ gives
\begin{align}
\label{eq:sintermsofe}
\hat {\mathfrak s} \simeq A \epsilon^{3a/2}
\end{align}
where  $A$ is a constant of integration.
But for this type of a localized black hole with three-charges in five
spacetime dimensions, we expect to find \cite{Strominger:1996sh}
\begin{align}
\label{eq:stoconst}
\hat {\mathfrak s} \to \textrm{constant} \ne 0
\end{align}
as $\epsilon\to0$.  This can only be true if $a=0$.
We have therefore seen that in order for the entropy of the small
three-charge black hole to be non-vanishing in the extremal limit,
the number $a$ in the ansatz \eqref{eq:ansatzepsilon} should
be zero. This is what makes the three-charge case special compared to the
one- and two-charge case.
The fact that $a$ vanishes has an immediate
consequence for the relative tension
\begin{align}
r = \frac{L \bar {\mathcal T}_z}{E} = \frac{\mu n}{\epsilon} =
\frac{n}{a+bn} = \frac{1}{b}.
\end{align}
{}From Equation (\ref{eq:erra}) we see that the five-dimensional
near-extremal three-charge black hole indeed has $a=0$ and $b=1/2$
which gives the correct value $r=2$.

Let us finally note that we can quickly see how $r$ depends on the
number of transverse dimensions and charges.\footnote{We choose
a short derivation here. One can also obtain the result by calculating
$\bar c_t$, $\bar c_z$, etc.} Firstly, in this general case we still
have that $L \bar {\cal T}_z=L {\cal T}_z=\mu n$ (see
Appendix~\ref{app:elec}). Thus we only have to see how $a$ and $b$
depends on $d$, the number of transverse spatial dimensions, and
$N_{\mathrm{ch}}$, the number of charges. We assume
that~\eqref{eq:chemicalpotential} applies and that the form of the
gauge fields is the same in the general case such that $\nu_a=\tanh
\al_a$ and
\begin{align}
 M^\textrm{el} = \sum_{a=1}^{N_{\mathrm{ch}}}\tanh
\al_a Q_a.
\end{align}
Further, from~\cite{Harmark:2004ws} we get $c_t\propto
\frac{(d-2)M-L{\cal T}}{(d-2)^2-1}$. Using this and the form of the
gauge fields, we see that
$Q_a=(d-3)\sinh\al_a\cosh\al_a\frac{(d-2)M-L{\cal T}}{(d-2)^2-1}$
thus giving
\begin{equation}
    \eps=\left(1-\frac{N_{\mathrm{ch}}(d-2)}{2(d-1)}\right)\mu
    +\frac{N_{\mathrm{ch}}}{2(d-1)}\mu n
\end{equation}
where we have used that $(\tanh\al_a-1)\sinh\al_a\cosh\al_a
\to-1/2$ as $\alpha_a \to \infty$. This, of course, agrees with our case where
$N_{\mathrm{ch}}=3$ and $d=4$. The only other case with $a=0$ is
for $N_{\mathrm{ch}}=4$ and $d=3$. For the latter case it is
actually known that one can have configurations with finite
entropy (see e.g.~\cite{Kallosh:1996ru}). However, our derivation
does not hold in this case since the asymptotic $c_t$ and $c_z$ do
not make sense for $d=3$.

\section{Phase diagrams for the near-extremal case}
\label{sec:3nearphase}

In this section we discuss consequences of the near-extremal
map for the different phases of the seeding solution considered in
Section \ref{sec:appl} and display the phase diagrams.

\subsection{Energy versus relative tension}
\label{sec:evsr}

In normal situations it would be appropriate to draw the different
phases of near-extremal solutions on an ($\epsilon, r)$ phase
diagram. But in the special case of a five-dimensional three-charge
black holes on a circle the relative tension $r$ is a
constant independent of the seeding solution. The phase diagram is
therefore just a straight line $r(\epsilon)=2$ which does not contain much
information about the different phases.

\begin{figure}
\begin{center}
\includegraphics[width=0.4\columnwidth]{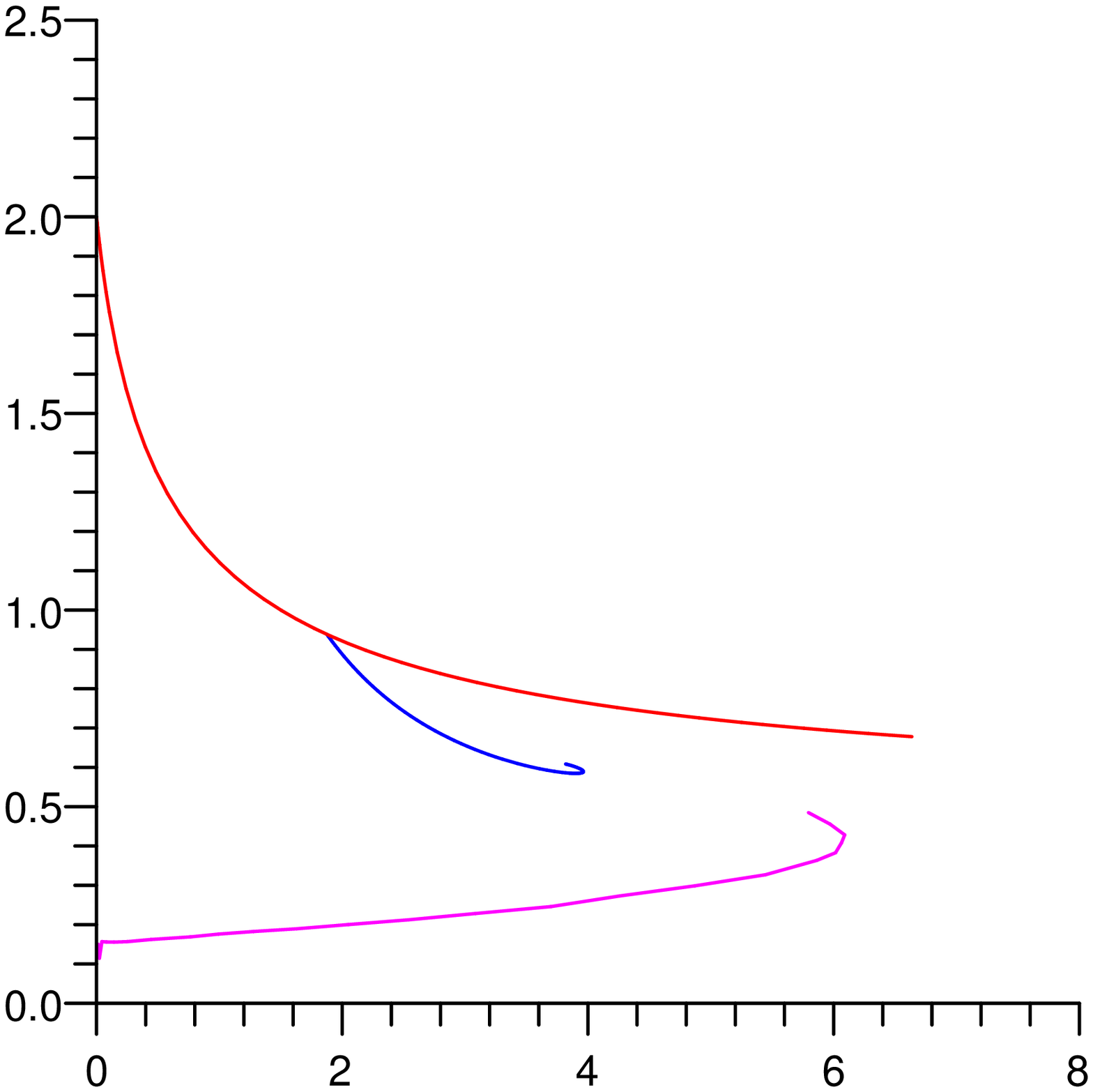}
\includegraphics[width=0.4\columnwidth]{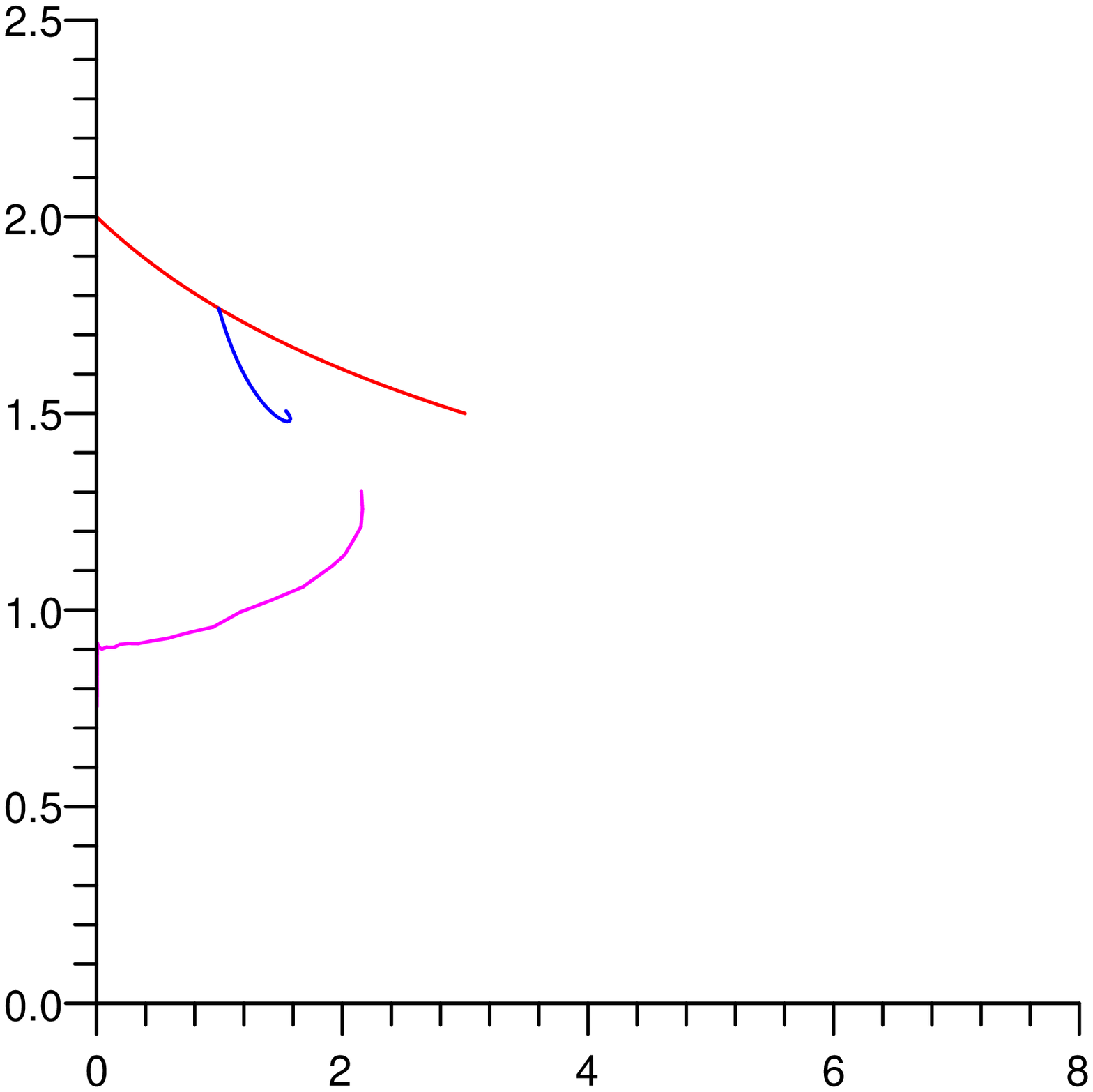}
\includegraphics[width=0.4\columnwidth]{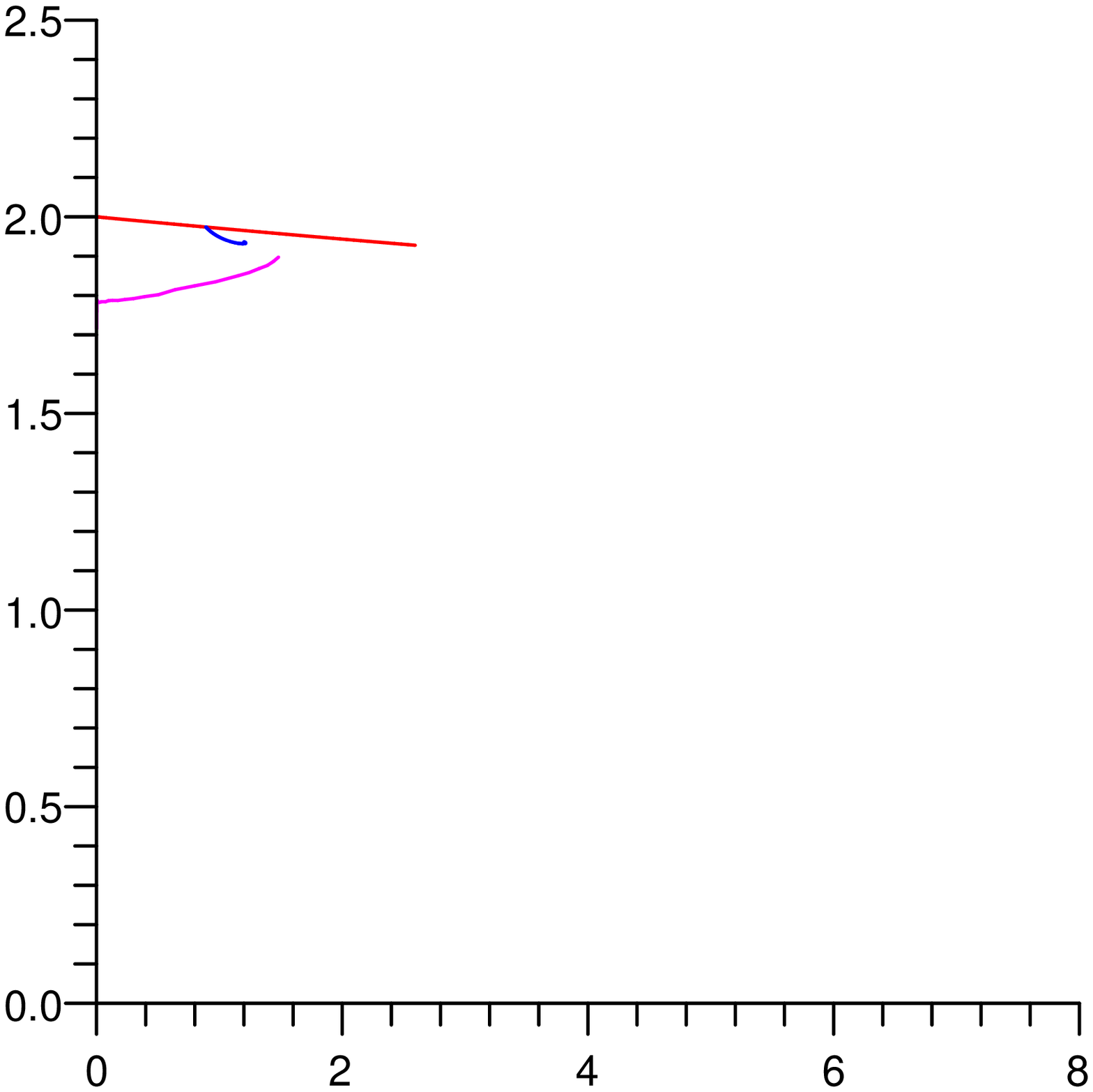}
\includegraphics[width=0.4\columnwidth]{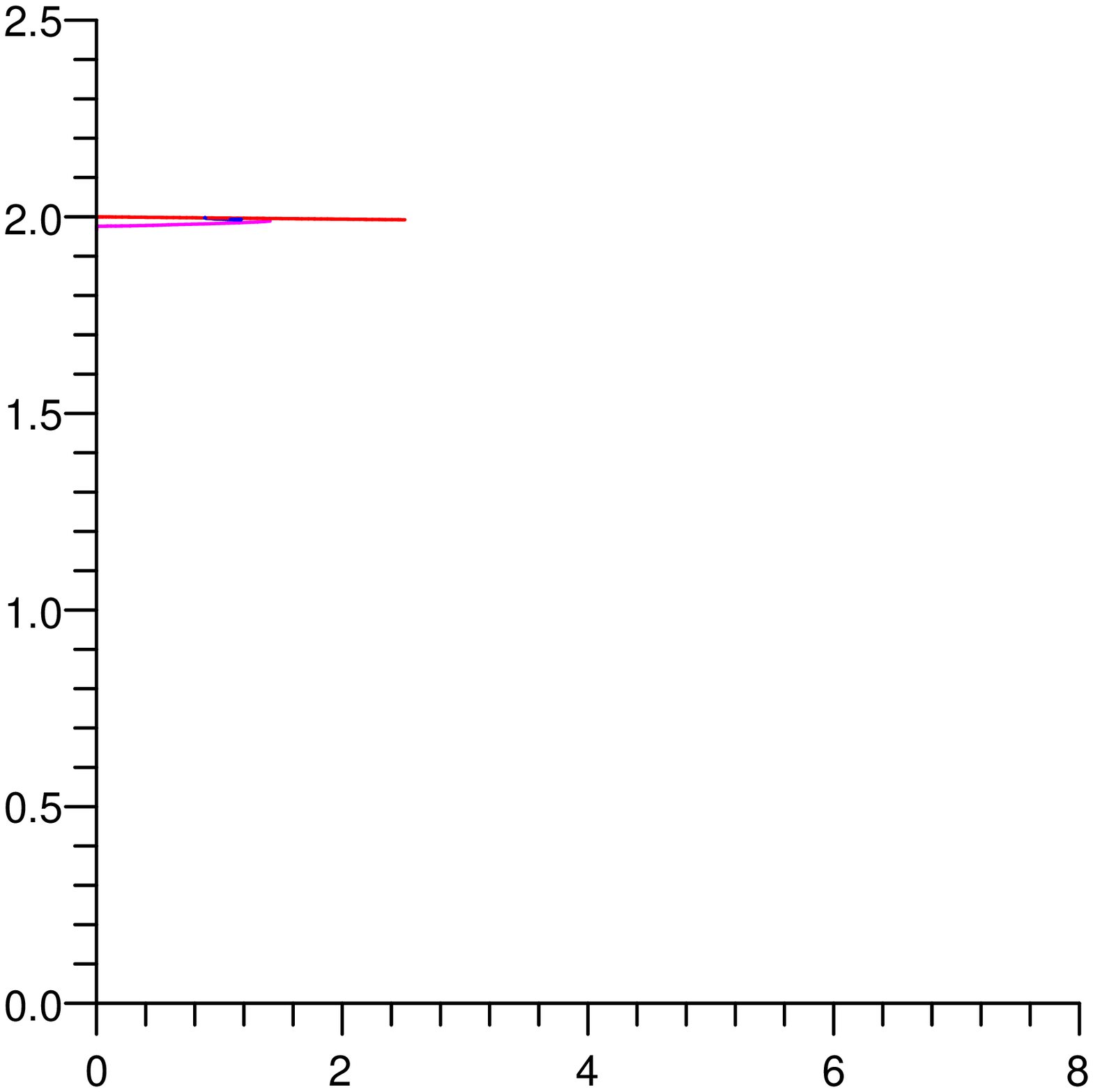}
\begin{picture}(0,0)(0,0)
\put(-180,22){\large $\eps$}\put(-343,159){\large $r$}
\put(-6,22){\large $\eps$}\put(-168,159){\large $r$}
\put(-6,194){\large $\eps$}\put(-168,331){\large $r$}
\put(-180,194){\large $\eps$}\put(-343,331){\large $r$}
\put(-230,320){\large $q=1$}\put(-55,320){\large $q=10$}
\put(-230,148){\large $q=100$}\put(-55,148){\large $q=1000$}
\end{picture}
\caption{The non-extremal $(\epsilon,r)$ phase diagram for four
different values of the charges. The three charges are all taken to
be equal and have the value $q=1$, $q=10$, $q=100$, and $q=1000$.
Notice how all the phases collapse to the line $r=2$ as the charges
go to infinity. The curves were found from Equation
(\ref{eq:goestor2}) using $n=1/2$ for the uniform phase (red curve),
numerical data from~\cite{Kudoh:2004hs}
for the localized phase (magenta curve) and numerical data
from~\cite{Kleihaus:2006ee} for the non-uniform phase (blue curve). }
\label{fig:rvse}
\end{center}
\end{figure}

We can, however, see how the relative tension approaches this constant
as the charges are sent to infinity.  In this discussion we
define, with a slight abuse of notation, the {\em non-extremal} energy above
extremality for finite charges as $\epsilon = \bar\mu -\sum_a q_a$.
{}From the definition of the relative tension, we then have that
\begin{align}
\label{eq:nonextremalr}
r = \frac{L \bar {\mathcal T}_z}{E} = \frac{\mu n}{\epsilon}.
\end{align}
We can plug in our equations for the tension and energy above
extremality \eqref{eq:map} and get
\begin{align}
\label{eq:goestor2}
\frac{\mu n}{\bar\mu - \sum_a q_a}
    &=  \left(\frac{1}{2} + \frac{(2-n)}{6n}\sum_a\frac{b_a}{1+\sqrt{1+b_a^2}}\right)^{-1}.
\end{align}
This quantity clearly goes to $r=2$ in the near-extremal limit,
since by Equation~\eqref{eq:ba} the $b_a$ vanish when the charges
go to infinity. We do not have analytic expressions for the full
localized phase nor for the non-uniform phase, but from the
numerical data \cite{Kudoh:2004hs,Kleihaus:2006ee}
 we can plot a non-extremal $(\epsilon,r)$ phase
diagram and see how it evolves as the charges go to infinity. Figure
\ref{fig:rvse} depicts this phase diagram for four increasing values
of the charges and we clearly see how all the phases collapse to the
degenerate line $r=2$ as the charges go to infinity.

\subsection{Thermodynamics of the uniform and non-uniform phases}
\label{sec:nearuniformthermo}

The thermodynamics of the uniform phase in the near-extremal limit
follows directly from the general map \eqref{eq:erra}, \eqref{eq:thatintermsof}
and the known thermodynamics ($\ms_{\rm u} (\mu) = \mu^2/4$)
of the uniform black string in five dimension, and we find
\begin{align}
\hat {\mathfrak s}_{\rm u}(\epsilon) = \sqrt{2\epsilon},
\label{eq:uniformF}
\qquad
\hat {\mathfrak f}_{\rm u}(\hat {\mathfrak t}) = -\frac{1}{2}\hat {\mathfrak t}^2.
\end{align}

If we apply the general map \eqref{eq:erra} to the neutral non-uniform
branch that was reviewed in Section \ref{sec:nuniphases} we get
a new non-uniform phase of near-extremal three-charge black holes
on a circle.  The Gregory-Laflamme point $(\mu_\textrm{GL},1/2)$
where the non-uniform phase branches off the uniform phase is mapped
to a critical point with energy above extremality
$\epsilon_\textrm{c} = \mu_\textrm{GL}/4$.
The relative tension at this point is $r=2$ as for all other points and therefore
we cannot describe the non-uniform phase as a curve on the $(\epsilon,r)$
diagram like in the non-extremal case.  We can, however, express the
neutral tension $n$ in terms of $\epsilon -\epsilon_\textrm{c}$ near the
critical point. The expression is
\begin{align}
\label{eq:nonunin}
n(\epsilon) = \frac{1}{2}  - \hat\gamma (\epsilon -\epsilon_\textrm{c})
+ \mathcal{O}((\epsilon -\epsilon_\textrm{c})^2)
\end{align}
with $\hat\gamma$ given by
\begin{align}
\label{eq:neargamma}
\hat\gamma = \frac{4\gamma}{1-2\gamma\mu_\textrm{GL}}
    = 38.89
\end{align}
where $\gamma = 0.14$ is the slope of the neutral non-uniform branch
and $\mu_\textrm{GL} = 3.52$ is the Gregory-Laflamme critical mass.
It is useful to have the neutral tension in terms of the energy
above extremality because we can integrate \eqref{eq:dlogsdloge} to
find the entropy for the non-uniform branch to leading order
\begin{align}
\label{eq:nonuniformS} \hat {\mathfrak s}_{\rm nu} (\epsilon) = \hat {\mathfrak
s}_\textrm{c}
    \left(1+ \frac{\epsilon -\epsilon_\textrm{c}}{2\epsilon_\textrm{c}}
    -\big(\frac{1}{8}+\frac{2}{3}\hat\gamma\epsilon_\textrm{c}\big)
    \frac{(\epsilon -\epsilon_\textrm{c})^2}{\epsilon_\textrm{c}^2}
    \right)
    +\mathcal{O}((\epsilon -\epsilon_\textrm{c})^3)
\end{align}
where $\hat {\mathfrak s}_\textrm{c} = \sqrt{2\epsilon_\textrm{c}}$ is the
critical entropy.

\begin{figure}
\begin{center}
\includegraphics[width=0.4\columnwidth]{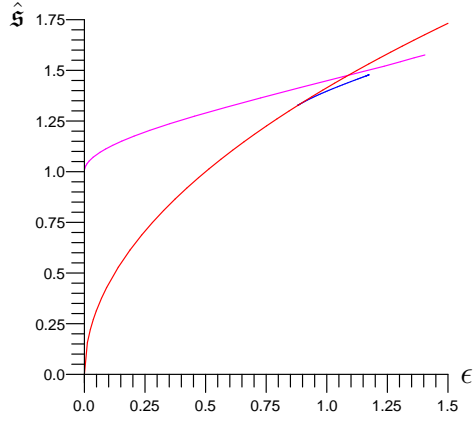}
\begin{picture}(0,0)(0,0)
\put(-6,26){\large $\epsilon$}\put(-176,161){\large $\hat {\mathfrak
s}$}
\end{picture}
\caption{The entropy $\hat {\mathfrak s}$ as a function of the
energy above extremality $\epsilon$ for the localized phase
(magenta), the uniform phase (red) and the non-uniform phase (blue).
The curves are based on numerical data
from~\cite{Kudoh:2004hs,Kleihaus:2006ee}.} \label{fig:svse}
\end{center}
\end{figure}

We can recover the entropy of the uniform branch by replacing
$\hat\gamma$ with zero in the expression \eqref{eq:nonuniformS}
above. Notice that the entropy of the non-uniform phase deviates
from that of the uniform phase only to second order.  These two
phases
\footnote{Note that for each of these two phases we also have copies, which are mapped
from the copies \cite{Horowitz:2002dc,Harmark:2003eg,Harmark:2004ws}
of the non-uniform and localized phase of the seeding solution. The thermodynamic quantities
of the copies of the near-extremal three-charge solutions are
given by $ \tilde \epsilon = \epsilon /k$, $\tilde{\hmt} = \hmt/\sqrt{k}$, $\tilde{\hms} = \hms/\sqrt{k}$ where $k = 2,3,\ldots$.}
are depicted in Figure \ref{fig:svse} together with the
localized phase which will be discussed in Section
\ref{sec:nearsmallthermo}.

\begin{figure}
\begin{center}
\includegraphics[width=0.4\columnwidth]{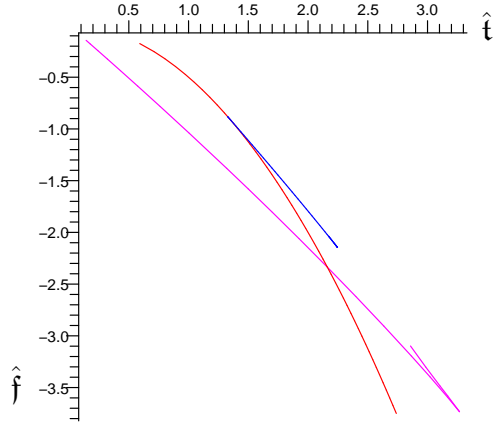}
\begin{picture}(0,0)(0,0)
\put(-1,145){\large $\hat {\mathfrak t}$}\put(-179,10){\large $\hat
{\mathfrak f}$}
\end{picture}
\caption{The free energy $\hat {\mathfrak f}$ as a function of the
temperature $\hat {\mathfrak t}$ for the localized phase (magenta),
the uniform phase (red) and the non-uniform phase (blue). The curves
are based on numerical data
from~\cite{Kudoh:2004hs,Kleihaus:2006ee}.} \label{fig:fvst}
\end{center}
\end{figure}

In the canonical ensemble we can get the free energy of the non-uniform
phase as an expansion around the critical temperature
$\hat {\mathfrak t}_\textrm{c} = \sqrt{2\epsilon_\textrm{c}}$.
Using the Smarr formula to relate temperature to energy above extremality,
we get from \eqref{eq:nonuniformS}
\begin{align}
\label{eq:nonuniformF} \hat {\mathfrak f}_{\rm nu} =
-\epsilon_\textrm{c}
    - \hat {\mathfrak s}_\textrm{c} (\hat {\mathfrak t}-\hat {\mathfrak t}_\textrm{c} )
    - \frac{c}{2\hat {\mathfrak t}_\textrm{c} }
        (\hat {\mathfrak t}-\hat {\mathfrak t}_\textrm{c} )^2
    + \mathcal{O}((\hat {\mathfrak t}-\hat {\mathfrak t}_\textrm{c} )^3)
\end{align}
where
\begin{align}
\label{eq:nonuniheatcap}
c= \frac{3\hat {\mathfrak s}_\textrm{c}}{3+16\hat\gamma\epsilon_\textrm{c}}
    = 0.0072
\end{align}
is the heat capacity of the non-uniform phase at $\hat {\mathfrak t}
= \hat {\mathfrak t}_\textrm{c}$. The free energy of the uniform
branch around $\hat {\mathfrak t} = \hat {\mathfrak t}_\textrm{c}$
is also given by \eqref{eq:nonuniformF} but with heat capacity
$c=\hat {\mathfrak s}_\textrm{c}$ as can be easily derived from
\eqref{eq:uniformF}. These phases are depicted in Figure
\ref{fig:fvst} together with the localized branch which will be
discussed in Section \ref{sec:nearsmallthermo}.

\subsection{Thermodynamics of small three-charge black holes on a circle}
\label{sec:nearsmallthermo}

We now consider the case of a small localized black hole. In the
neutral case we have (see Section~\ref{sec:locphase})
\begin{align}
\label{eq:munrho} \mu = \frac{3}{2}\rho_0^2 + \frac{1}{32}\rho_0^4 +
\mathcal{O}(\rho_0^8), \quad n = \frac{1}{24}\rho_0^2 -
\frac{1}{1152}\rho_0^4 + \mathcal{O}(\rho_0^6)
\end{align}
which gets mapped by \eqref{eq:erra} to the energy above extremality
\begin{align}
\label{eq:epsrho}
\epsilon  =  \frac{1}{32}\rho_0^4 + \mathcal{O}(\rho_0^8).
\end{align}
Note that not only is the $\rho_0^2$ term missing but the $\rho_0^6$
term cancels as well. The rescaled entropy and temperature are
by~\eqref{eq:thatintermsof} mapped into
\begin{align}
\label{eq:shatrho} \hat{\mathfrak s}   &=
    1+ \frac{\rho_0^2}{16} + \frac{\rho_0^4}{512} + {\mathcal O}(\rho_0^6),\\
\label{eq:thatrho} \hat{\mathfrak t} &=
    \rho_0^2 - \frac{\rho_0^4}{16} + \frac{\rho_0^6}{512} + {\mathcal O}(\rho_0^8).
\end{align}
Using~\eqref{eq:epsrho} we can also write the entropy in terms of
$\eps$
\begin{align}
\label{eq:smallentropy} \hat{\mathfrak s}_{\rm loc} (\epsilon) =
1+\sqrt{\frac{\epsilon}{8}} + \frac{\epsilon}{16}
    + \mathcal{O}(\epsilon^{3/2}).
\end{align}
showing the first two corrections to the extremal entropy for a
thermal black hole localized on a circle.  We correctly see that the
entropy \eqref{eq:smallentropy} goes to a non-vanishing constant
in the extremal limit $\epsilon \to 0$.
It is not surprising that the first correction comes with a power of
$\epsilon$ smaller than one, otherwise the temperature would not go
to zero in the extremal limit.

Restoring the normalization of the
entropy using~\eqref{eq:lllg} we find
\begin{align}
\label{eq:bigShatrho} \hat S_{\rm loc}  &= \frac{\ell \hat{\mathfrak
s}}{g}  = 2\pi \sqrt{N_1N_4N_0}
    \left(1+\sqrt{\frac{\epsilon}{8}} + \frac{\epsilon}{16}
    + \mathcal{O}(\epsilon^{3/2})\right).
\end{align}
The entropy in the extremal limit is the constant $\hat
S_0 = 2\pi \sqrt{N_1 N_4 N_0}$ in agreement with the well-known result of
\cite{Strominger:1996sh}. Eq.~\eqref{eq:bigShatrho} is one of the central
results of the paper and gives, as a function
of the energy above extremality, the first two corrections to the finite entropy
due to the interactions of the black hole. We will come back to this in Section
\ref{sec:microstates}, where we will present a microscopic counting of the
corrected entropy in case of the partial extremal limit described in Section \ref{sec:fincha}.

The Helmholtz free energy of small localized black holes in the
canonical ensemble \eqref{eq:freeenergy} is given by
\begin{align}
\label{eq:3nearfreesmall} \hat{\mathfrak f}_{\rm loc}(\hat{\mathfrak
t})
    = -\hat{\mathfrak t}
    - \frac{1}{32}\hat{\mathfrak t}^2
    - \frac{1}{512}\hat{\mathfrak t}^3
    + \mathcal{O}(\hat{\mathfrak t}^4).
\end{align}
The fact that the leading term in the free energy is linear in
$\hat{\mathfrak t}$ is in accord with the localized black hole
background being asymptotically $AdS_2\times S^3\times T^5$.
One expects the dual gauge theory to be quantum mechanical
and hence the free energy to be proportional to the temperature.
The higher order terms are then due to the presence of the circle.

It is also interesting to see how the thermodynamics changes if only
one or two of the charges are sent to infinity with the others kept
finite. These cases are studied in the next section.

\section{Other near-extremal limits}
\label{sec:extensions}

In this section we discuss some other limiting cases for the
near-extremal three-charge background, involving one or two finite
charges. We also present  the special case of
non-extremal solutions with two charges only and present the corresponding
near-extremal limit.

\subsection{Near-extremal limit with finite charges \label{sec:fincha}}

We start by  considering near-extremal limits where one or two of
the three charges stay finite.  This corresponds to having one or
more of the $\gamma_a$ in \eqref{eq:nearextremallimit} vanishing.

\subsubsection*{One finite charge}

Without loss of generality we can choose any of the three charges
finite. We choose here to take $q_4, q_0 \to\infty$ with $q_1$
finite. The presence of this finite charge corresponds to
choosing $\gamma_1 = 0$ with $\gamma_4, \gamma_0 > 0$ in the
near-extremal limit \eqref{eq:nearextremallimit}. The explicit form
of the resulting background is easily obtained using the general expressions in
\eqref{eq:nearextremalmetric}--\eqref{eq:Aafork}. This limit is also called the
dilute gas limit in the literature \cite{Horowitz:1996fn} and after a
T-duality in the $x$-direction where the F1-string lies, corresponds to the
near-extremal D1-D5 brane system with finite KK momentum in the
direction of the D-string.

In close analogy to \eqref{eq:nearextremalET}, the energy and tensions in this
partial limit are defined as
\begin{align}
\label{Elim1} E & = \lim_{L\to0} \left(\bar M - \sum_{a=0,4}
Q_a\right), \quad \hat{\mathcal T}_z = \lim_{L\to0}
\frac{L}{2\pi}\bar {\mathcal T}_z, \cr \quad \hat{\mathcal T}_1 & =
\lim_{L\to0}  (L_1\bar {\mathcal T}_1 - L_1\bar {\mathcal T}_1^{\rm el}) , \quad L_a \hat{\mathcal T}_a
= \lim_{L\to0} \left( L_a\bar {\mathcal T}_a -Q_a\right) \textrm{ for } a =0,4.
\end{align}
The dimensionless versions of these quantities are taken to be
\begin{align}
\label{deflim1} \epsilon = gE , \quad \quad r = \frac{2\pi \hat
{\mathcal T_z}}{E-M_1^{\rm el}}, \quad \quad r_a = \frac{L_a \hat
{\mathcal T}_a}{E-M_1^{\rm el}} \spa \hmt = \ell \hat T \spa \hms
= \frac{g}{\ell} \hat S
\end{align}
where $M_1^{\rm el}$ is the electric mass defined in \eqref{eq:Mela}, $g$ is defined
in \eqref{eq:gdefined} and $\ell = \ell_0 \ell_4$.

Using the definitions \eqref{Elim1}, \eqref{deflim1} and the results
in \eqref{eq:barM}--\eqref{eq:L0T0} for the physical quantities of the general three-charge
background, one finds after some algebra
\begin{align}
\label{maplim1} \epsilon = \frac{1 +n}{3} \mu +  \mu_1^{\rm el} \spa
r= \frac{3n}{1+n} \spa r_1 = 1 \spa r_a = \frac{3n}{2(1+n)} ,
\quad\textrm{for }a=4,0,
\end{align}
where we used Equations \eqref{eq:mundefined} to
write the final result in terms of the physical parameters $\mu$ and
$n$ of the original seeding black hole. This provides for this
partial near-extremal case the map from the neutral solution to the
charged one. In this case, the only relative tension that is
constant is the one in the spatial world-volume direction corresponding
to the charge that is finite.

For temperature and entropy one easily finds the mapping
\begin{equation}
\label{tslim1} \hmt = \frac{\mt^2 \ms}{\cosh \al_1} \spa \hms =
\mt^{-1} \cosh \al_1.
\end{equation}
We also recall that $\mu_1^{\rm el} = \nu_1 q_1$, with the chemical
potential $\nu_1$ and charge $q_1$ given by \eqref{eq:chempot},
\eqref{eq:qintermsof}  in terms of $\al_1$.
Finally, the Smarr relation in this case takes the form
\begin{align}
\label{eq:2chargeSmarr} \hmt \hms = (2-r)\left( \epsilon -
\mu^\textrm{el}_1 \right) .
\end{align}

The above map can of course be applied in particular to the neutral
solutions that fall into the black hole/string ansatz
\eqref{eq:metansatz3ch}, as was done in Section \ref{sec:3nearphase}
for the full near-extremal limit. For later use, we present here the
result for the localized phase, obtained by applying the map
\eqref{maplim1} to the localized black hole on a circle. The
corrected background for the near-extremal two-charge localized
black hole follows by taking the near-extremal limit
\eqref{eq:nearextremallimit} of the non-extremal background
\eqref{eq:metansatz3chtilde} with the appropriate choice of
$\gamma_a$. In particular, this amounts to $H_{a} \rightarrow \hat
H_a$ where $\hat H_a$ are given in \eqref{eq:Hafork}.

The thermodynamic quantities of the resulting near-extremal
localized phase, carrying one finite charge $q_1$ are then given by
\begin{align}\label{eq:thermoonefinite}
\epsilon & =  \rho_0^2\sinh^2\alpha_1
    + \frac{1}{2} \rho_0^2 \left( 1 + \frac{1}{16} \rho_0^2\right)
    + {\cal{O}} (\rho_0^6), \\
r &=  \frac{1}{8} \rho_0^2 - \frac{1}{128} \rho_0^4 + {\mathcal O}(\rho_0^6),\\
\hms &= \rho_0 \cosh \al_1 \left(1 + \frac{1}{16} \rho_0^2 + \frac{1}{512} \rho_0^4
\right) + {\mathcal O}(\rho_0^7),
 \\
\hmt & = \frac{\rho_0}{\cosh \al_1} \left(1- \frac{1}{16} \rho_0^2 + \frac{1}{512} \rho_0^4 \right) + {\mathcal O}(\rho_0^7).
\end{align}

In Section \ref{sec:microstates} we will provide a microscopic derivation
of the entropy found in the case of one finite charge. However, we will
consider a permutated version of the above limit, namely
with the D0-brane charge kept finite and F1 and D4-brane charge sent to infinity.
It is not difficult to see that this case is completely analogous to the one discussed
above.

In these expressions, we can send $\alpha_1\to 0$ (and hence $q_1 \rightarrow 0$)
and obtain the entropy and temperature of a small localized two-charge black hole
on a circle (see Section \ref{sec:2charge}). The entropy clearly vanishes
in the extremal limit $\rho_0\to0$.

\subsubsection*{Two finite charges}

We now choose $q_4  \to\infty$ with $q_1$ and $q_0$ finite,
corresponding to taking $\gamma_4 =1 $ with  $\gamma_1 = \gamma_0=
0$ in the expressions \eqref{eq:nearextremallimit}. Again, the explicit form of the
resulting background is easily obtained using the general
expressions in \eqref{eq:nearextremalmetric}--\eqref{eq:Aafork}.

The energy and tensions in this partial limit are defined by the
obvious generalizations of \eqref{Elim1} and the dimensionless
quantities are similar to those in \eqref{deflim1}, where we now
divide by $E - M_1^{\rm el} - M_0^{\rm el}$.

We then find after some algebra the map
\begin{equation}
\epsilon = \frac{4 +n}{6} \mu + \mu_1^{\rm el} + \mu_0^{\rm el},
\end{equation}
\begin{align}
r = \frac{6n}{4+n} \spa r_a  = \frac{2(1+n)}{4+n} ,
\quad\textrm{for } a=1,0 \spa r_4 = \frac{3n}{4+n},
\end{align}
where we recall that $\mu_a^{\rm el} = \nu_a q_a$, with the chemical
potential $\nu_a$ and charge $q_a$ given by \eqref{eq:chempot},
\eqref{eq:qintermsof}  in terms of $\al_a$.
For vanishing $q_1$ and $q_0$ the above results agree with the
one-charge $d=4$ case considered in \cite{Harmark:2004ws}.

For temperature and entropy one easily finds the mapping
\begin{equation}
\label{tslimm1} \hmt = \frac{\mt^{3/2} \ms^{1/2} }{\cosh \al_1 \cosh
\al_0} \spa \hms = \mt^{-1/2} \ms^{1/2} \cosh \al_1 \cosh \al_0
\end{equation}
Finally, the Smarr relation in this case takes the form
\begin{align}
\label{eq:2chargeSmarr2} \hmt \hms = \frac{1}{2}(2-r)\left( \epsilon
- \mu^\textrm{el}_1 - \mu^\textrm{el}_0 \right) .
\end{align}

As before, all of this can be applied to the ansatz, and in
particular for the localized phase we now get
\begin{align}\label{eq:thermoonefinite2}
\epsilon & =  \rho_0^2(\sinh^2\alpha_1 + \sinh^2\alpha_0)
    + \rho_0^2 \left( 1 + \frac{1}{32} \rho_0^2 \right)
    + {\cal{O}} (\rho_0^6),\\
 r &=  \frac{1}{16} \rho_0^2 - \frac{1}{512} \rho_0^4 + {\mathcal O}(\rho_0^6), \\
\hms &= \rho_0^2 \cosh \al_1 \cosh \al_0
    \left(1 + \frac{1}{16} \rho_0^2 + \frac{1}{512} \rho_0^4 \right)
    + {\mathcal O}(\rho_0^8),
 \\
\hmt & = \frac{1}{\cosh \al_1 \cosh \al_0 }
\left(1- \frac{1}{16} \rho_0^2 + \frac{1}{512} \rho_0^4 \right) + {\mathcal O}(\rho_0^6).
\end{align}
 The entropy vanishes in the extremal
limit, as we expect, but one finds finite extremal temperature. This
is in accord with the fact that we know that for $d=4$ (see e.g.
\cite{Harmark:2004ws}) the localized phase corresponds to the
near-extremal Type II NS5-brane, which has a Hagedorn temperature.

\subsection{Two-charge black holes on a circle \label{sec:2charge} }

Starting with the general three-charge non-extremal case one may
also consider the situation with a smaller number of non-zero
charges. In this section, we present some further details for the case with
two non-zero charges, which was not studied before.

\subsubsection*{Brief review of one-charge case}

As remarked earlier, when we set two of the three charges
equal to zero, we should recover the one-charge case which was
extensively studied in Ref.~\cite{Harmark:2004ws,Harmark:2005pq}. In particular,
by sending say $q_1,q_0 \to 0$, Eq.~\eqref{eq:map} becomes
\begin{align}
\bar \mu - q_4 &= \frac{(4+n)\mu}{6}
    + \frac{(2-n)\mu }{6} \frac{b_4}{1+\sqrt{1+b_4^2}}
\end{align}
which agrees with Eq.~(4.18) of \cite{Harmark:2004ws} for $d=4$.
Recall that $b_a$ was defined in Equation \eqref{eq:ba}. As
an example, for the localized phase discussed in Section \ref{sec:locphase}
we can eliminate $\mu$ and $n$ to arrive at \cite{Harmark:2004ws}
\begin{align}
\label{loc1ch} \bar n(\bar \mu; q_4) = \frac{1}{24} (\bar\mu- q_4)
    + {\mathcal O}\left((\bar\mu - q_4)^2\right).
\end{align}
For comparison below, we also give here the map from the neutral
five-dimensional Kaluza-Klein black holes to the near-extremal
one-charge physical quantities
\begin{equation}
\label{eq:map1}
\epsilon = \frac{4 +n}{6} \mu   \spa r = \frac{6 n}{4 +n} \spa r_4 = \frac{3n}{4+n}
 \spa
\hmt =  \mt^{3/2} \ms^{1/2} \spa \hms = \mt^{-1/2} \ms^{1/2}.
\end{equation}

\subsubsection*{Non-extremal two-charge case}

Turning to the two-charge case,
we keep $q_0$, $q_4$ finite and take $q_1 \to 0$. The non-extremal background can
simply be obtained by setting $\al_1 =0$ in the general form in
\eqref{eq:einsteinmetric}--\eqref{eq:Ha}.
For the thermodynamic quantities, we can use for example
\eqref{eq:map} to compute the non-extremal map
\begin{align}
\label{nonex2charge} \bar \mu - \sum_{a=0,4}q_a &=
\frac{(1+n)\mu}{3}
          + \frac{(2-n)\mu }{6} \sum_{a=0,4} \frac{b_a}{1+\sqrt{1+b_a^2}}
\end{align}
and for the temperature and entropy we simply have \eqref{eq:bart},
\eqref{eq:bars} with $\al_1 =0$.

As an application of \eqref{nonex2charge}, it follows using the
results of Section \ref{sec:locphase} that  for the localized phase
\begin{align}
\label{loc2ch} \bar n(\bar \mu; q_4,q_0) = \frac{1}{12} \left(\bar
\mu - q_4-q_0\right)
       +  {\mathcal O}\left((\bar \mu - q_4 -q_0)^2\right).
\end{align}
We observe from \eqref{loc1ch} and \eqref{loc2ch} that
 in both the two-charge and the one-charge case, the relative
tension for a small black hole is linear in the energy above
extremality. For the three-charge case, which is special in many
respects, this is not the case (see  Eq.~\eqref{eq:map}).

\subsubsection*{Near-extremal limit}

We consider here the near-extremal limit of the two-charge black
hole solution, in which we send both of the charges to infinity.
Before discussing these results we briefly review the near-horizon
limit of the corresponding extremal two-charge background.

For the localized phase, corresponding to D0-D4 smeared in the $x$-direction,
we find after a T-duality in that direction the D1-D5 brane system, which
has near-horizon geometry $AdS_3 \times S^3 \times T^4$.
As a consequence we expect that the leading behavior of the thermodynamics of
the localized phase in the near-extremal two-charge system
with a transverse circle corresponds to that of a two-dimensional CFT.
As we will see below this is indeed the case.
For the uniform phase we find that the extremal background is described by a
doubly-smeared configuration of D0-D4 branes, which after a double T-duality
(in the $x$ and $z$-direction) corresponds to the D2-D6 brane system. The dual
description of this is less clear and presumably gravity is not decoupled, due
to the presence of the D6-brane. However, as we will see below the system
exhibits a Hagedorn behavior, in close analogy to the near-extremal NS5-brane system
 \cite{Maldacena:1996ya,Maldacena:1997cg}.

The definition of the near-extremal two-charge limit follows the same route as discussed in
Section \ref{sec:3chargenearextr} for the three-charge case, where now the harmonic
function $H_1$ is set to one. The corresponding background follows
likewise from \eqref{eq:nearextremalmetric}--\eqref{eq:Hafork} by taking $\gamma_0 > 0$, $\gamma_4 >0$ and
setting $\hat H_1 =1$. In this case the quantity $\ell $ (of dimension length)
that enters the dimensionful physical quantities is given by
\begin{equation}
\label{eq:ell2ch}
\ell = \ell_0 \ell_4  = \frac{ (2\pi l_s)^5 g_s }{L V_1 \sqrt{V_4}} \sqrt{N_0 N_4}
\end{equation}
where we used \eqref{eq:ell42}, \eqref{eq:ell02}.
The quantity $g$ is still given by \eqref{eq:ggstring}.

The physical quantities are defined as in \eqref{eq:nearextremalET}
with $\al_1 =0$ (and hence $M_1^{\rm el} = Q_1 = 0$) and the
dimensionless quantities  are as in \eqref{eq:neardimless} (in this
case there is no ${\cal{T}}_1$). The map follows easily by setting
$q_1 =0$ in the map \eqref{maplim1}, \eqref{tslim1} so that the map
from neutral Kaluza-Klein black holes to near-extremal two-charge
physical quantities is
\begin{equation}
\label{eq:map2}
\epsilon = \frac{1 +n}{3} \mu   \spa r = \frac{3 n}{1 +n} \spa r_a =
\frac{3n}{2(1+n)} \textrm{ for } a =4,0 \, ,
\end{equation}
\begin{equation}
\hmt =  \mt^2 \ms \spa \hms = \mt^{-1}.
\end{equation}
Like the one-charge map in \eqref{eq:map1}, this is a one-to-one map that maps from the
neutral two-dimensional $(\mu,n)$ phase diagram  to the
two-dimensional $(\epsilon,r)$ phase diagram of near-extremal
two-charge solutions.

The Smarr relation is
\begin{equation}
\hmt \hms  = (2-r) \epsilon .
\end{equation}
It is useful to recall that for a given curve in the $(\epsilon,r)$
phase diagram we can find the entire thermodynamics from this Smarr
relation and the first law of thermodynamics $\delta \epsilon = \hmt
\delta \hms$, by integrating the equation
\begin{equation}
\frac{ \delta \log \hms (\epsilon)  }{ \delta \log \epsilon} =
\frac{1}{2 - r(\epsilon)} \, .
\end{equation}

In particular, for the solutions that are generated from the ansatz
\eqref{ansatz} the metric takes the form of
\eqref{eq:nearextremalmetric}--\eqref{eq:Aafork} with $\hat H_1 =1$. For
the three known phases of black holes/strings on a cylinder discussed in
Section \ref{sec:ansatz} we can
then map to the corresponding phases of two-charge black holes with a
circle in the transverse space.

For this we can use, as in Section \ref{sec:appl}, the known data for the
phases of five-dimensional Kaluza-Klein black holes along with the
analytically known results for the uniform phase, the non-uniform
phase near the GL point and localized phase in the small mass limit.%
\footnote{Note that the non-uniform and localized phases also have copies, which are mapped
from the copies \cite{Horowitz:2002dc,Harmark:2003eg,Harmark:2004ws}
of the non-uniform and localized phase of the seeding solution. The thermodynamic quantities
of the copies of the near-extremal two-charge solutions are
given by $ \tilde \epsilon = \epsilon /k$, $\tilde{\hmt} = \hmt$, $\tilde{\hms} = \hms/k$ where $k = 2,3,\ldots$.}

\begin{figure}
\begin{center}
\includegraphics[width=0.4\columnwidth]{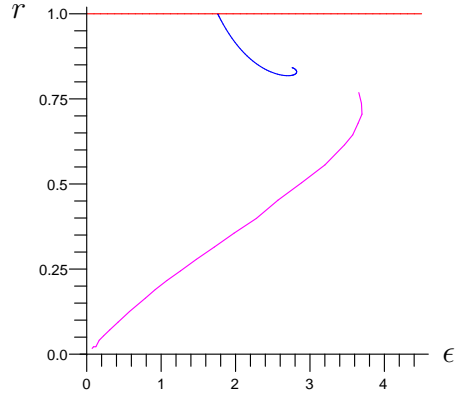}
\begin{picture}(0,0)(0,0)
\put(-6,26){\large $\epsilon$}\put(-169,156){\large $r$}
\end{picture}
\caption{$ (\epsilon,r)$ phase diagram for near-extremal two-charge
black holes on a circle. Shown are the localized phase (magenta),
the uniform phase (red) and the non-uniform phase (blue). The curves
are based on numerical data
from~\cite{Kudoh:2004hs,Kleihaus:2006ee}.}
 \label{fig:epsr}
\end{center}
\end{figure}

\begin{figure}
\begin{center}
\includegraphics[width=0.4\columnwidth]{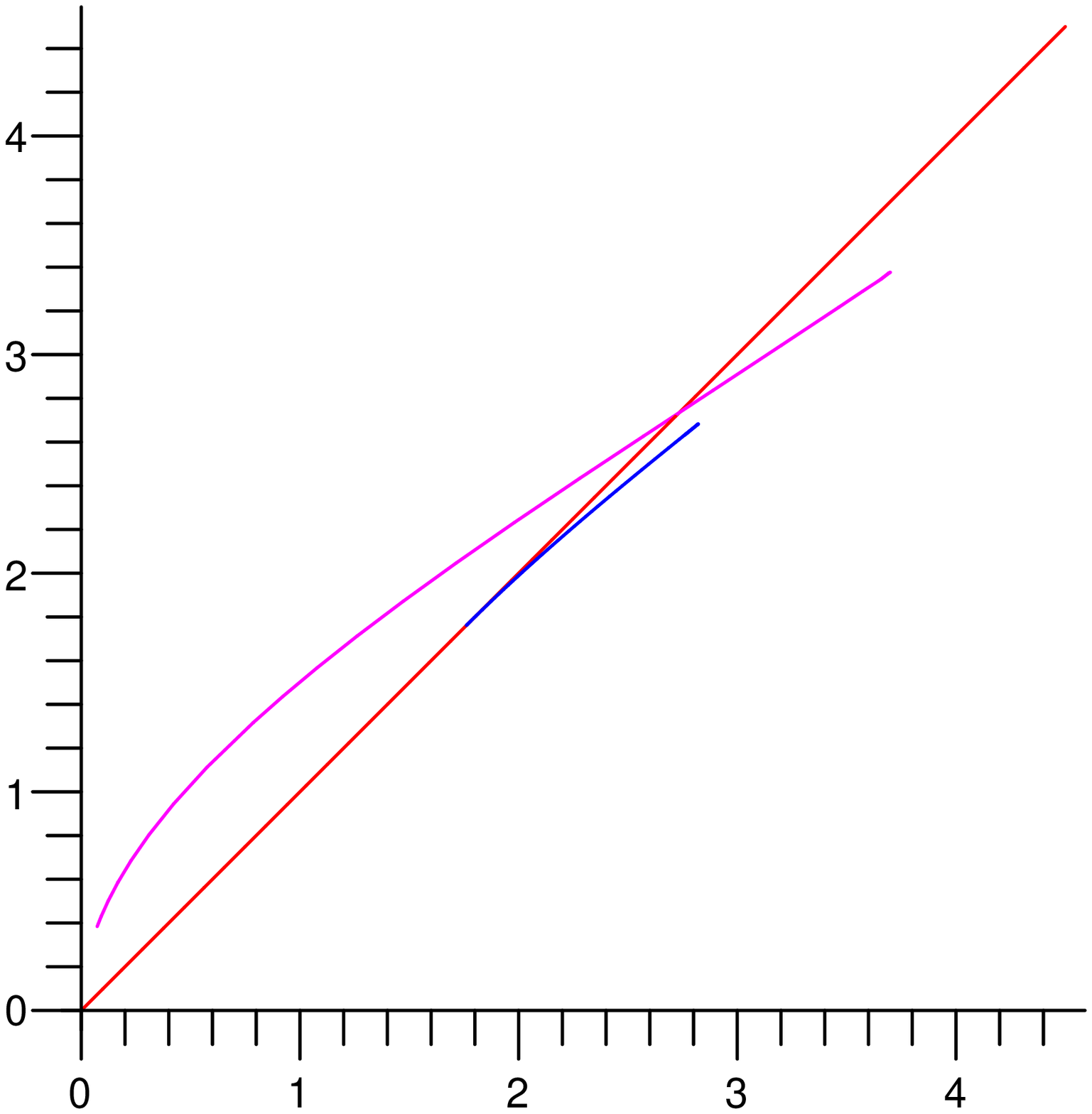}
\hspace{5pt}
\includegraphics[width=0.4\columnwidth]{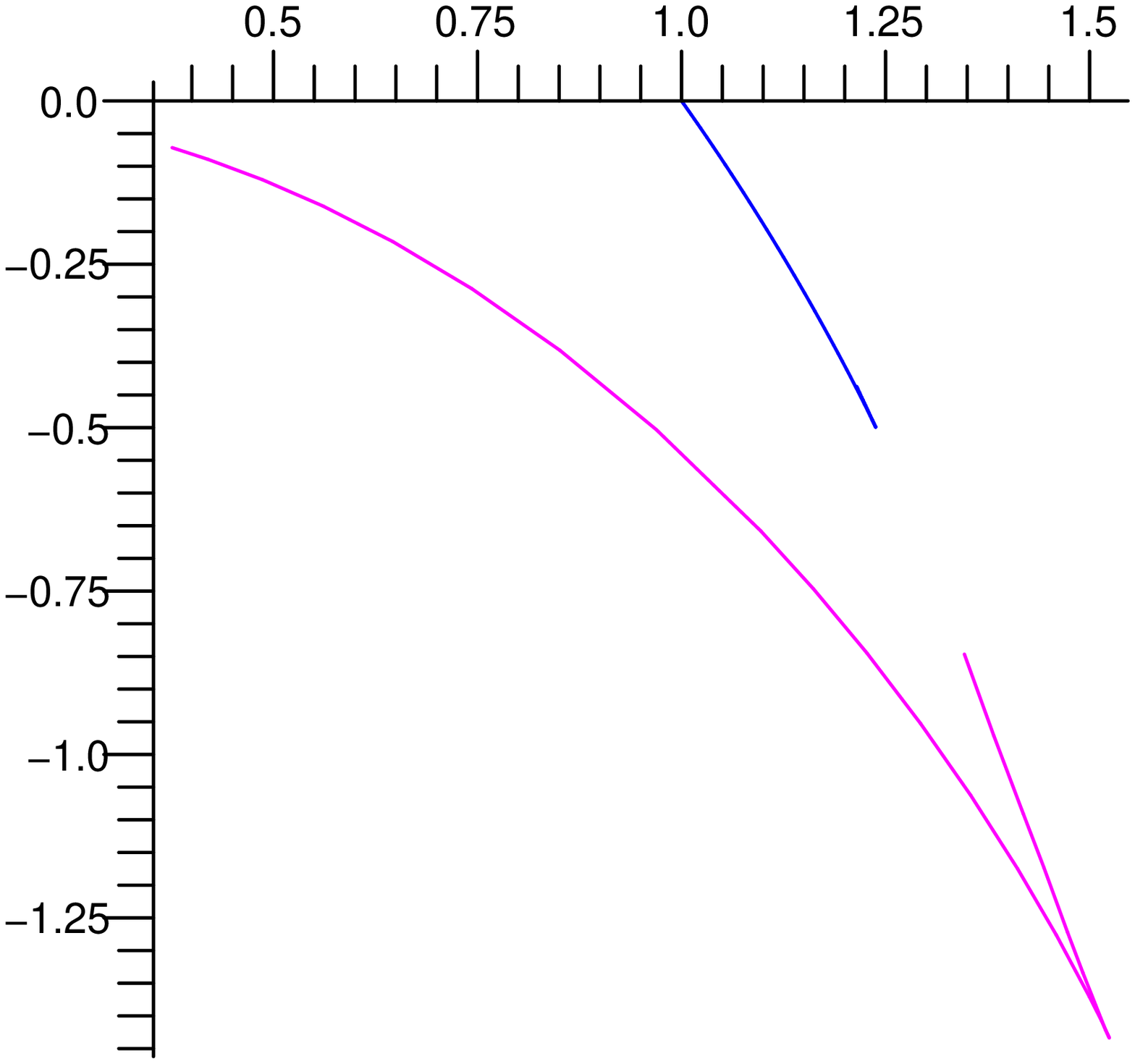}
\begin{picture}(0,0)(0,0)
\put(-192,26){\large $\epsilon$}\put(-352,159){\large $\hat
{\mathfrak s}$}\put(-3,137){\large $\hat {\mathfrak
t}$}\put(-165,10){\large $\hat {\mathfrak f}$}
\end{picture}
\caption{$(\epsilon,\hms)$ and $(\hmt,\mathfrak{f})$ diagrams for
near-extremal two-charge black holes on a circle. Shown are the
localized phase (magenta), the uniform phase (red) and the
non-uniform phase (blue). The curves are based on numerical data
from~\cite{Kudoh:2004hs,Kleihaus:2006ee}. } \label{fig:esps2}
\end{center}
\end{figure}

The results can be summarized are as follows. For the uniform phase
we have
\begin{equation}
r_{\rm u} (\epsilon) = 1
\spa
\hms_{\rm u} (\epsilon) = \epsilon
\spa
\hat{\mathfrak{f}}_{\rm u} (\hmt) = 0
\end{equation}
showing that this phase has Hagedorn thermodynamics with Hagedorn
temperature $\hmt_{\rm c} =1$ found e.g. from $t^{-1}=\partial
s/\partial \epsilon$. For the non-uniform phase we have
\begin{equation}
r_{\rm nu} (\epsilon) = 1 - \hat \gamma \cdot ( \epsilon -
\epsilon_{\rm c})  + {\cal{O}}\left( ( \epsilon -
\epsilon_{\rm c})^2 \right),
\end{equation}
\begin{equation}
\hms_{\rm nu} (\epsilon) =  \hms_{\rm u} (\epsilon) \left(1  -
\frac{\hat \gamma}{2 \epsilon_{\rm c}} ( \epsilon - \epsilon_{\rm
c})^2 + {\cal{O}}\left( ( \epsilon -
\epsilon_{\rm c})^3 \right)  \right),
\end{equation}
 \begin{equation}
\hat{\mathfrak{f}}_{\rm nu}  (\hmt) = - \epsilon_{\rm c} (\hmt -1) -
\frac{1}{2 \hat \gamma} (\hmt -1)^2  + {\cal{O}}\left( ( \hmt -
1)^3 \right)
\end{equation}
\begin{equation}
\epsilon_{\rm c} = \frac{\mu_{\rm GL}}{2} = 1.76 \spa \hat \gamma =
\frac{8 \gamma}{3 - 2  \mu_{\rm GL} \gamma} = 0.56,
\end{equation}
exhibiting the departure at the critical point from the Hagedorn
thermodynamics.
Finally, for the localized phase
\begin{equation}
r_{\rm loc} (\epsilon) =  \frac{1}{4} \epsilon - \frac{1}{16}
\epsilon^2 + {\cal{O}} ( \epsilon^3 ),
\end{equation}
\begin{equation}
\label{sloc2}
\hms_{\rm loc} (\epsilon) = \sqrt{2} \epsilon^{1/2} \left(
1 + \frac{1}{16} \epsilon- \frac{1}{512} \epsilon^2 +  {\cal{O}} ( \epsilon^3 ) \right),
\end{equation}
\begin{equation}
\label{floc2}
\hat{\mathfrak{f}}_{\rm loc} (\hmt) = -\frac{1}{2} \hmt^2 - \frac{1}{32} \hmt^4
- \frac{1}{256} \hmt^6  + {\cal{O}} ( \hmt^8 ).
\end{equation}
The numerically obtained plots for all of these quantities are shown
in Figures \ref{fig:epsr} and \ref{fig:esps2}.

We first observe that the leading order thermodynamics (for small
temperatures) of the localized phase of the near-extremal D1-D5
brane system on a transverse circle correctly exhibits a free energy
that is proportional to $\hmt^2$, as expected for a two-dimensional
conformal field theory at finite temperature. The results in
\eqref{sloc2}, \eqref{floc2} then describe the departure of this
behavior due to higher order temperature corrections in the presence
of the circle.

On the other hand, note that the near-extremal uniformly smeared D1-D5 brane phase
exhibits a Hagedorn temperature $T_{\rm hg} = 1/\ell$ (since $\hmt_{\rm c} =1$)
with $\ell$ given in \eqref{eq:ell2ch}, with
the non-uniformly smeared phase emerging at the Hagedorn temperature. The picture
that emerges in this system is in many respects analogous to the one
considered in Ref.~\cite{Harmark:2005dt} where the thermodynamics of near-extremal NS5-branes was
studied, and applied to Little String Theory. In particular, we see from the above
that also here, the localized phase provides a new stable phase
in the canonical ensemble, extending to a maximum temperature that lies above
the Hagedorn temperature. It would be interesting
to see if there is a dual interpretation of this, which would require
a further examination of the near-extremal D2-D6 brane system.

\section{Microscopic entropy}
\label{sec:microstates}

In this section we use the microstate counting technique of
\cite{Strominger:1996sh, Horowitz:1996ay} to recalculate the entropy
of our three-charge black holes on a circle. The non-extremal branes
will interact across the transverse circle and this interaction
effectively shifts the number of branes \cite{Costa:2000kf}. In the
case of a small localized near-extremal black hole with one finite
charge, we find agreement between the first correction obtained via
microstate counting and the macroscopic corrected entropy in
Equation (\ref{eq:thermoonefinite}).

\subsection{Review of non-extremal black hole microstate counting}
\label{sec:microreview}

Horowitz, Maldacena and Strominger \cite{Horowitz:1996ay} showed how
to count the microstates for a special class of five-dimensional
non-extremal black holes with three charges. In their case the
metric asymptotes to Minkowski space with no circle in the
transverse space.

In the weak string coupling limit the extremal black hole can be
described as a configuration with $N_4$ D4-branes, $N_1$ fundamental
strings and $N_0$ D0-branes. Non-extremal black holes can be
generated, for example, by adding a small number $N_{\bar 0}$ of
anti-D0-branes. In a thermal system one would of course expect the
non-extremality also to excite the D4-branes and the F1-strings, but
we assume that the anti-D0-brane excitation is much lighter than the
other two.
The mass of the black hole is then given by the sum of the masses of
each type of object
\begin{align}
\label{eq:micromass} M = V_1\tau_1 N_1 + V_4\tau_4 N_4 + V_0 \tau_0
\left( N_0 + N_{\bar 0}\right)
\end{align}
where $V_a$ is the world-volume of each object, $\tau_1 = (2\pi
\ell_s^2)^{-1}$ is the tension of the string, $\tau_4 = (g_s
(2\pi)^4 \ell_s^5)^{-1}$ is the tension of the D4-brane and $\tau_0
= (g_s \ell_s)^{-1}$ is the tension of the D0-brane. The charge
associated to the D0-branes is given by
\begin{align}
\label{eq:microcharge} Q_0 = V_0 \tau_0 \left( N_0 - N_{\bar
0}\right)
\end{align}
while the other charges are extremal and therefore given by $Q_a =
V_a \tau_a N_a$ for $a=1,4$.

The D4-branes are separated in the spatial world-volume direction of
the F1-string. There are therefore effectively $N_4N_1$ strings
between neighboring D4-branes. The D0-branes are like beads
threaded on any one of these strings and in the dilute gas limit the
strings are far apart and the beads can therefore only be threaded
by one string at a time. In the extremal case with $N_{\bar 0}=0$,
this gives rise to an entropy \cite{Horowitz:1996ay}
\begin{align}
\label{eq:extremalentropy} S = 2\pi \sqrt{N_1N_4N_0}.
\end{align}
In the non-extremal case the anti-D0-branes are like beads with
opposite charge. In the dilute  gas limit the forces between the
beads are small and so interactions can be ignored.  The entropy is
additive in this case and given by \cite{Horowitz:1996ay}
\begin{align}
\label{eq:entropynbarn} S = 2\pi \sqrt{N_1N_4}\left( \sqrt{N_0} +
\sqrt{N_{\bar 0}} \right).
\end{align}
This is the equation that we want to generalize for our
near-extremal three-charge black hole with one finite charge
on a circle.

\subsection{Microstate counting on a circle}
\label{sec:microcircle}

In the previous subsection the beads were far apart and did not
interact, but with the small transverse circle present that is no
longer a safe assumption to make.
In our near-extremal limit, the size of the transverse circle was
taken to be at the same scale as the energy above extremality and
the interaction energy is therefore not negligible compared to the
excitation energy.  That means that interactions across the
transverse circle between beads of opposite charge must be taken
into account.
The effect of the interaction is to shift the number of beads for a
given total energy \cite{Costa:2000kf}.  We now examine this
for the localized phase of three-charge black holes on a circle.%
\footnote{We thank Roberto Emparan for suggesting this computation to us.}

The non-extremal mass of the three-charge black hole \eqref{eq:barM}
can be rewritten as
\begin{align}
\label{eq:microbarM} \bar M = \frac{\Omega_2 c_t}{2gL} \left[ \left(
1- 2\frac{c_z}{c_t}\right)
    + \cosh2\alpha_1+ \cosh2\alpha_4 + \cosh2\alpha_0 \right].
\end{align}
The first term is equal to
\begin{align}
\label{eq:firstterm} \tilde E \equiv \frac{\Omega_2 c_t}{2gL} \left(
1- 2\frac{c_z}{c_t}\right)
\end{align}
and we notice that $\tilde E$ is proportional to the tension along the
transverse circle.  This term is absent in the case where there is
no transverse circle. The terms involving $\cosh2\alpha_a$ are
recognized as the contribution of each type of extended object to
the total mass of five-dimensional three-charge black hole without
the circle \cite{Horowitz:1996ay}.  It is therefore natural to write
\begin{align}
\label{eq:barMsplit} \bar M = \tilde E + \bar M_1 + \bar M_4 + \bar
M_0
\end{align}
with
\begin{align}
\label{eq:barMdoubleangle} \bar M_a = \frac{\Omega_2 c_t}{2gL}
\cosh2\alpha_a.
\end{align}
The charges \eqref{eq:qa} can also be rewritten as
\begin{align}
\label{eq:Qadoubleangle} Q_a = \frac{\Omega_2 c_t}{2gL} \sinh
2\alpha_a.
\end{align}
It is now easy to see that in the full near-extremal limit we have
$\bar M_a - Q_a \to 0$ and $\bar M-\sum_a Q_a  \to \tilde E$.

\subsubsection*{Partial extremal limit}

Let us now consider the case where two of the charges are taken to
be extremal, say $Q_1$ and $Q_4$, while $Q_0$ has some small
non-extremality.  The total mass of the black hole is then
\begin{align}
\label{eq:splittM} \bar M = Q_1 + Q_4 + \bar M_0 + \tilde E.
\end{align}
Following Costa and Perry \cite{Costa:2000kf}\footnote{The same idea
has been applied in \cite{Emparan:2001bb}.}, we want to write the
total mass in the form
\begin{align}
\label{eq:deltaEVint} \bar M = Q_1 + Q_4 + \delta E + V_\textrm{int}
\end{align}
where $\delta E$ is the energy carried by the D0-branes and the
anti-D0-branes, and $V_\textrm{int}$ is the interaction energy
related to the presence of the transverse circle.  As we start
adding anti-D0-branes to the extremal system, they will interact
with the D0-branes across the circle, reducing their energy by
$V_\textrm{int}$.

The force between the beads across the circle gives rise to the
tension $\mathcal{T}$ and the interaction energy is the ``energy
stored in the tension".  In $d=4$ the tension is proportional%
\footnote{Since ${\cal{T}} = n M/L $ and $M \propto L^{d-2}$
\cite{Harmark:2002tr} we have ${\cal{T}} \propto L^{d-3}$.}
to $L$ and therefore
\begin{align}
\label{eq:Vint} V_\textrm{int} = - \int \mathcal{T} dL =
-\frac{1}{2} \mathcal{T} L.
\end{align}
Notice that from the near-extremal map \eqref{eq:erra} we have
$\frac{1}{2} \mathcal{T} L  = \tilde E$ and hence $V_\textrm{int} =
- \tilde E$.

Equation \eqref{eq:splittM} can now be written as
\begin{align}
\label{eq:barMVint} \bar M = Q_1 + Q_4 + (\bar M_0 +2\tilde E) +
V_\textrm{int}
\end{align}
so we can identify
\begin{align}
\label{eq:deltaE} \delta E = \bar M_0 + 2\tilde E.
\end{align}

We find the effective number of D0- and anti-D0-branes from
requiring
\begin{align}
\label{eq:deltaENbarN}
\delta E &= V_0 \tau_0 (N'_0 + N'_{\bar 0}), \\
\label{eq:QNbarN} Q_0 &= V_0\tau_0( N'_0 - N'_{\bar 0}).
\end{align}
This gives
\begin{align}
\label{eq:N0} \tau_0 N'_0 &= \frac{1}{2}\left( \bar M_0 + Q_0\right)
+ \tilde E
       = \frac{1}{2} \frac{\Omega_2 c_t}{2gL} \exp({2\alpha_0}) + \tilde E, \\
\label{eq:barN0} \tau_0 N'_{\bar 0} &=  \frac{1}{2}\left( \bar M_0 -
Q_0\right) + \tilde E
       = \frac{1}{2} \frac{\Omega_2 c_t}{2gL} \exp({-2\alpha_0}) + \tilde E
\end{align}
where we used the expression \eqref{eq:Qadoubleangle} for $Q_0$. We thus see
that there is a shift of $\tilde E$ in the effective
number of zero-branes compared to the black hole without the
transverse circle \cite{Horowitz:1996ay,Costa:2000kf}.  Recall that
$\tau_0 = 1/g_s\ell_s$ and $V_0=1$.

The microstate entropy for our interacting system on a circle is
then given by
\begin{align}
\label{eq:Soursystem} S = 2\pi \sqrt{N_1 N_4} \left( \sqrt{N'_0} +
\sqrt{N'_{\bar 0}}\right)
\end{align}
where $N'_0$ and $N'_{\bar 0}$ are the effective number of D0- and
anti-D0-branes given in Equations \eqref{eq:N0}--\eqref{eq:barN0}.

\subsubsection*{Application to small localized three-charge black holes}

For small (neutral) localized black holes we have
$c_t = \rho_0^2L /(4\pi)$ and recall from Eq.~\eqref{eq:chisqrtA} that
$c_z/c_t = \chi = 1/2 - \rho_0^2/32 + \mathcal{O}(\rho_0^6) $.
We therefore get that
\begin{align}
\label{eq:smallexample} \frac{\Omega_2 c_t}{2L} =
\frac{1}{2}\rho_0^2, \qquad  g\tilde E = \frac{1}{32} \rho_0^4 +
\mathcal{O}(\rho_0^6)
\end{align}
where we used the definition of $\tilde E$ in \eqref{eq:firstterm}.
We thus compute from \eqref{eq:N0}, \eqref{eq:barN0} the expressions
\begin{align}
\label{eq:obviousnot} g\tau_0 N'_{0} &= \frac{1}{4} \rho_0^2
\exp(2\alpha_0)
        + \frac{1}{32} \rho_0^4+\mathcal{O}(\rho_0^6),\\
g\tau_0 N'_{\bar 0} &= \frac{1}{4} \rho_0^2 \exp(- 2\alpha_0)
        + \frac{1}{32} \rho_0^4+\mathcal{O}(\rho_0^6).
\end{align}
To this order we therefore get
\begin{align}
\label{eq:sqrtnsqrtnbar} \sqrt{N'_{0}} + \sqrt{N'_{\bar 0}}
    = \frac{\sqrt{N_0}}{\ell_0}
    \rho_0 \cosh\alpha_0 \left(1 + \frac{\rho_0^2}{16}+\mathcal{O}(\rho_0^4)\right)
\end{align}
where we have used Equation \eqref{eq:ell02} to rewrite $g\tau_0 =
\ell_0^2/N_0$. The microstate entropy in Equation
\eqref{eq:Soursystem} then becomes
\begin{align}
\label{eq:ourmicroS} S &= \frac{2\pi \sqrt{N_1 N_4N_0}}{\ell_0}
    \rho_0 \cosh\alpha_0
        \left(1+ \frac{\rho_0^2}{16}+\mathcal{O}(\rho_0^4)\right)\\
    &= \frac{\ell_1\ell_4}{g}  \rho_0 \cosh\alpha_0
        \left(1+ \frac{\rho_0^2}{16}+\mathcal{O}(\rho_0^4)\right).
\end{align}
This agrees with our previous result for the partial near-extremal
entropy obtained from the black hole
side~\eqref{eq:thermoonefinite}, up to the order $\rho_0^4$ term.

We could in principle include higher order terms in Equations
\eqref{eq:smallexample} and hope to find agreement in the entropy to
higher order, but it is not clear to which degree the method of
shifting the effective number of branes is accurate. We do not
expect the microstate picture to hold for the uniform or the
non-uniform phase so it is clear that somewhere on the way it must
break down.

{}From Equation \eqref{eq:epsrho} we know that the $\rho_0^6$ order
term in $\tilde E$ is vanishing and this information would yield
$(1-2\cosh\alpha_0)\rho_0^4/512$ as the next order term in the
parenthesis of Equation \eqref{eq:ourmicroS}.  This is to be
compared to $\rho_0^4/512$ from the Bekenstein-Hawking entropy on
the black hole side. It is not too surprising to find a minor
discrepancy at such an high order.

\section{Conclusions \label{sec:concl} }

In this paper we have generated new three-charge black hole
solutions for the situation in which the three-charge black holes
have a transverse circle. As we have shown, this case is interesting
since the presence of the circle has a non-trivial effect for the
physics of the three-charge system, giving rise to several new
phenomena. Of particular interest is the case of a small
three-charge black hole localized on the circle. Here we are
deformed away from the non-zero entropy $S=2\pi\sqrt{N_1 N_4 N_0}$
of the extremal case.

The starting point of our work is the construction of a map from
 neutral black hole solutions in five-dimensional
Kaluza-Klein space-time to the three-charge black holes. This map
is obtained using boosts and U-dualities, and is an extension of
the map for the one-charge case obtained in Ref.~\cite{Harmark:2004ws}. We
restricted ourselves to solutions without Kaluza-Klein bubbles,
and hence the phases we found were the localized phase, where the
three-charge black is localized on the circle, the uniform phase,
where the three-charge black hole is smeared uniformly around the
circle, and the non-uniform phase, where the three-charge black
hole is smeared without gaps on the circle in a translationally
non-invariant fashion. As a consequence of the mapping of phases
we also find an ansatz for three-charge black holes on a circle
which applies to all of the considered new phases.

After constructing the non-extremal solution we turned to the
near-extremal limit of the solutions. Here we have several new
phenomena appearing that make the three-charge case special in
comparison to the one-charge case of \cite{Harmark:2004ws}. One of
our results is that the relative tension always is constant $r=2$,
i.e.\ the tension along the circle is always proportional to the
energy with the same constant of proportionality, even for the
localized case when the energy is very small. We show that this is
not a coincidence but rather a hitherto unknown consequence of the
non-zero entropy of the extremal limit.

Having constructed the three-charge solutions, we examined their
physical properties, including the global thermodynamic stability
in both the microcanonical and canonical ensembles. We find the
corrections to the extremal black hole entropy for the case of the
small three-charge black hole localized on a circle. We also obtain the
corrections to the thermodynamics for the non-uniform phase when
it is close to the critical point. Finally, we examine how the
constant relative tension $r=2$ is approached as the near-extremal
limit $q \rightarrow \infty$ is taken.

We considered furthermore the two-charge solutions. One interesting feature
in this case is that the system exhibits a Hagedorn behavior, with the localized phase
providing a stable phase that extends up to a limiting temperature that lies above
the Hagedorn temperature. The same behavior was recently found in the thermodynamics
of the near-extremal NS5-brane \cite{Harmark:2005dt}. More generally,
the two-charge system has the interesting aspect that the AdS/CFT correspondence tells us
that our deformations of the usual two-charge solution without the
presence of the transverse circle should have a counterpart as
deformations of the dual two-dimensional CFT. This would be very
interesting to investigate.

Finally, we examined whether one can extend the microstate
counting of the entropy performed in  \cite{Horowitz:1996ay}  to the case of a small
three-charge black hole localized on the transverse circle. This
was done following a similar analysis as that of Ref.~\cite{Costa:2000kf}. Interestingly,
we find that in fact the first correction to the entropy due to
the presence of the transverse circle is in perfect agreement with
the microstate counting of the entropy. This first correction
takes into account the self-interaction of the black hole across
the circle, giving rise to a non-zero potential depending on the
circumference of the circle. It would be very interesting to see
if this matching of the black entropy and microstate counting
could work to higher order. However, this would seemingly require
a better understanding of the contributions to the potential for
the black hole, since at higher order effects like gravitational
backreaction can play a role.

There are several future directions for research on three-charge
black hole solutions with a transverse circle. One is to construct
more new solutions for three-charge black holes. Indeed, using the
map with the five-dimensional bubble-black hole sequences of
\cite{Elvang:2004iz} as neutral seeding solutions will produce new
three-charge solutions with regular event horizons. This could be very
interesting to examine, for one thing the free energy of the
near-extremal limit of such solutions can be negative, as one can
see from \eqref{eq:freemixed} and the fact that there exist plenty of
bubble-black hole configurations with $n \leq 4/5$. This could
very well hint at the existence of new stable phases of the
three-charge system.

Another interesting avenue to pursue would be to consider in more
detail the non-uniform phase and the stability of the uniform
phase. For both the two-charge and three-charge black holes we
found a new non-uniform phase emerging from the uniform phase
where the black hole is smeared along the transverse circle. That
we in this way can dress the non-uniform phase with charges has
in the one-charge case been shown
\cite{Aharony:2004ig,Harmark:2005jk} to be connected with a map from the
Gregory-Laflamme mode of the neutral black string to an unstable
mode of the singly-charged uniform phase. It would be interesting
to examine whether this works similarly for the two- and
three-charge black holes. This involves construction a map of the
unstable mode which in addition to boosts and U-dualities also
would include complex rotations as in \cite{Aharony:2004ig,Harmark:2005jk}.
Obviously this would be highly interesting in view of the
Correlated Stability Conjecture (CSC) of
\cite{Gubser:2000ec,Gubser:2000mm,Reall:2001ag,Gregory:2001bd}, which is
examined in \cite{Harmark:2005jk} for the one-charge case,
also in view of recent work on the CSC for other brane bound
states \cite{Gubser:2004dr,Ross:2005vh,Friess:2005tz}.

Since we successfully can match the black hole entropy and the
microstate counting for a small three-charge black hole localized
on a circle, it would be interesting to examine further the
consequences of this for Mathur's fuzzball proposal
\cite{Mathur:2005zp,Mathur:2005ai}. Mathur's proposal is to
reproduce the black hole entropy by counting smooth supergravity
solutions without an event horizon. It would be interesting if one
could reproduce our result for the entropy in this way.

\section*{Acknowledgments}

We thank Burkhard Kleihaus, Jutta Kunz and Eugen Radu for kindly
providing their data on the non-uniform branch  in five dimensions
and Hideaki Kudoh and Toby Wiseman for kindly providing their data
on the black hole branch in five dimensions. We also thank Shinji
Hirano and Vasilis Niarchos for useful discussions. Special thanks
to Roberto Emparan for illuminating remarks and discussions.
TH and NO would also like to thank the KITP for hospitality while
part of this work was completed.
 Work partially supported by the European Community's Human Potential
Programme under contract MRTN-CT-2004-005104 `Constituents,
fundamental forces and symmetries of the universe'.

\appendix

\section{The boosts and U-dualities}
\label{sec:Uduality}

In this appendix we show how the neutral solution is charged
up via boosts and U-dualities.
Let us start with a static and neutral five-dimensional Kaluza-Klein
black hole as a seeding solution.  The metric of such a solution can
be written in the form
\begin{equation}
ds_5^2 = -U dt^2 + \frac{L^2}{(2\pi)^2} V_{ab}dx^a dx^b
\end{equation}
where $V_{ab}dx^a dx^b$ describes a cylinder in the asymptotic
region of circumference $L$.
There is no dilaton and no gauge fields.
By adding flat dimensions $x$ and $u_i$, $i=1,...,4$, and performing a
series of boosts and U-dualities we can construct the ten-dimensional
solution of Type IIA Supergravity given in Section \ref{sec:adding3charges}.

\subsection{The route}

Before going through the calculation, let us first schematically
sketch the route that we will take.
First we make a boost in $t$ and $x$ direction with boost-parameter $\alpha_1$:
\begin{center}
\begin{tabular}{c|cccccc}
& $t$ & $x$ & $u_1$ & $u_2$ & $u_3$ & $u_4$  \\
\hline
($\alpha_1$) P & $\times$ & $\times$
\end{tabular}
\hspace{1cm} Type IIB
\end{center}
T-dualize in $x$ direction:
\begin{center}
\begin{tabular}{c|cccccc}
& $t$ & $x$ & $u_1$ & $u_2$ & $u_3$ & $u_4$  \\
\hline
($\alpha_1$) F1 & $\times$ & $\times$
\end{tabular}
\hspace{1cm} Type IIA
\end{center}
Boost in $t$ and $x$ direction with boost-parameter $\alpha_4$:
\begin{center}
\begin{tabular}{c|cccccc}
& $t$ & $x$ & $u_1$ & $u_2$ & $u_3$ & $u_4$  \\
\hline
($\alpha_1$) F1 & $\times$ & $\times$ \\
($\alpha_4$) P & $\times$ & $\times$
\end{tabular}
\hspace{1cm} Type IIA
\end{center}
Lift to M-theory by adding the 11th dimension $y$:
\begin{center}
\begin{tabular}{c|ccccccc}
& $t$ &$y$ & $x$ & $u_1$ & $u_2$ & $u_3$ & $u_4$  \\
\hline
($\alpha_1$) M2 & $\times$ &$\times$& $\times$ \\
($\alpha_4$) P & $\times$ && $\times$
\end{tabular}
\hspace{1cm} M-theory
\end{center}
Go back to Type IIA by reducing on $x$:
\begin{center}
\begin{tabular}{c|cccccc}
& $t$ &$y$  & $u_1$ & $u_2$ & $u_3$ & $u_4$  \\
\hline
($\alpha_1$) F1 & $\times$ &$\times$ \\
($\alpha_4$) D0 & $\times$
\end{tabular}
\hspace{1cm} Type IIA
\end{center}
T-dualize in $u_1,u_2,u_3,u_4$:
\begin{center}
\begin{tabular}{c|cccccc}
& $t$ &$y$  & $u_1$ & $u_2$ & $u_3$ & $u_4$  \\
\hline
($\alpha_1$) F1 & $\times$ &$\times$ \\
($\alpha_4$) D4 & $\times$ &&$\times$ &$\times$ &$\times$ &$\times$
\end{tabular}
\hspace{1cm} Type IIA
\end{center}
Lift to M-theory again by adding an 11th dimension $x$:
\begin{center}
\begin{tabular}{c|ccccccc}
& $t$ &$x$ &$y$  & $u_1$ & $u_2$ & $u_3$ & $u_4$  \\
\hline
($\alpha_1$) M2 & $\times$& $\times$ &$\times$ \\
($\alpha_4$) M5 & $\times$& $\times$ &&$\times$ &$\times$ &$\times$ &$\times$
\end{tabular}
\hspace{1cm} M-theory
\end{center}
Boost in $t$ and $x$ with boost parameter $\alpha_0$:
\begin{center}
\begin{tabular}{c|ccccccc}
& $t$ &$x$ &$y$  & $u_1$ & $u_2$ & $u_3$ & $u_4$  \\
\hline
($\alpha_1$) M2 & $\times$& $\times$ &$\times$ \\
($\alpha_4$) M5 & $\times$& $\times$ &&$\times$ &$\times$ &$\times$ &$\times$ \\
($\alpha_0$) W & $\times$& $\times$
\end{tabular}
\hspace{1cm} M-theory
\end{center}
Go back to IIA by reducing on $x$:
\begin{center}
\begin{tabular}{c|cccccc}
& $t$ &$y$  & $u_1$ & $u_2$ & $u_3$ & $u_4$  \\
\hline
($\alpha_1$) F1 & $\times$ &$\times$ \\
($\alpha_4$) D4 & $\times$& &$\times$ &$\times$ &$\times$ &$\times$ \\
($\alpha_0$) D0 & $\times$
\end{tabular}
\hspace{1cm} Type IIA
\end{center}

We stop here with a configuration that is a thermal excitation of an F1-string, D4-brane and a D0-brane, but we could T-dualize in directions $u_3$ and $u_4$ and lift to M-theory once more to get the configuration:
\begin{center}
\begin{tabular}{c|ccccccc}
& $t$ &$x$&$y$  & $u_1$ & $u_2$ & $u_3$ & $u_4$  \\
\hline
($\alpha_1$) M2 & $\times$ & $\times$ & $\times$ \\
($\alpha_4$) M2 & $\times$ & & & $\times$ &$\times$ \\
($\alpha_0$) M2 & $\times$ & & & & &$\times$&$\times$
\end{tabular}
\hspace{1cm} M-theory
\end{center}
This is a thermal excitation of a configuration that is known to be $1/8$-BPS.

\subsection{Transformation of the solution}

We now examine how the solution transforms under the boosts and U-dualities
described in the previous subsection.  We start with the metric
\begin{align}
ds^2_{10} = -U dt^2 + dx^2 + \sum_{i=1}^4 (du^i)^2 + ds_4^2
\end{align}
where we have introduced the shorthand
$ds_4^2 \equiv  \frac{L^2}{(2\pi)^2} V_{ab}dx^a dx^b$.
This is considered to be a solution of Type IIB Supergravity with vanishing
dilaton and no gauge fields present.

Under a Lorentz-boost along the $x$-axis with rapidity $\alpha_1$,
the coordinates transform as
\begin{align}
\left( \begin{array}{c} t_\textrm{new} \\ x_\textrm{new} \end{array} \right)
=
\left( \begin{array}{cc} \cosh\alpha_1 & \sinh\alpha_1 \\
                                   \sinh\alpha_1 & \cosh\alpha_1 \end{array} \right)
\left( \begin{array}{c} t_\textrm{old} \\ x_\textrm{old} \end{array} \right)
\end{align}
and the metric becomes
\begin{align}
ds^2_{10} = &\left(-U\cosh^2\alpha_1 + \sinh^2\alpha_1\right) dt^2
                  -  2(1-U) \cosh\alpha_1\sinh\alpha_1 dt dx \nonumber\\
                 &+ \left(-U\sinh^2\alpha_1 + \cosh^2\alpha_1\right)dx^2
                 + \sum_{i=1}^4 (du^i)^2 + ds_4^2.
\end{align}

There is an isometry in the $x$ direction and we can therefore use equations (2.54) in
\cite{Peet:2000hn} to T-dualize in that direction and get a solution of Type IIA Supergravity.
The dilaton becomes (fields with/without a tilde are new/old)
\begin{align}
e^{2\tilde \phi} = \frac{e^{\phi}}{g_{xx}} =
\frac{1}{\left(-U\sinh^2\alpha_1 + \cosh^2\alpha_1\right)} = H_1^{-1}
\end{align}
where we have defined
\begin{align}
H_1 \equiv \left(-U\sinh^2\alpha_1 + \cosh^2\alpha_1\right)
= 1 + (1-U)\sinh^2\alpha_1.
\end{align}
The components of the metric that change under the duality are
\begin{align}
\tilde g_{xx} &= \frac{1}{g_{xx}} = H_1^{-1} \\
\tilde g_{tx} &= 0\\
\tilde g_{tt} &= g_{tt} - (g_{tx})^2/g_{xx}= -U H_1^{-1}
\end{align}
and we get a Kalb-Ramond field with component
\begin{align}
\tilde B_{tx} = \frac{g_{tx}}{g_{xx}} = \coth\alpha_1 \left(H_1^{-1} - 1\right).
\end{align}
Therefore the solution has become
\begin{align}
ds^2_{10} &= H_1^{-1}\left( -U dt^2 + dx^2 \right) + \sum_{i=1}^4 (du^i)^2 + ds_4^2 \\
e^{2\phi} &= H_1^{-1} \\
B &= \coth\alpha_1 \left(H_1^{-1} - 1\right) dt \wedge dx
\end{align}
and we see that we have already picked up one charge.  Two to go.

To produce the second charge we make another Lorentz boost in the $x$
direction, now with boost parameter $\alpha_4$.  The effect on the metric
is analogous to the one above, except now all terms with $dt$ and $dx$ are
multiplied with $H_1^{-1}$. The dilaton is a scalar and does therefore not
transform and it turns out that B is also invariant because
\begin{align}
dt_\textrm{new} \wedge dx_\textrm{new} = dt_\textrm{old} \wedge dx_\textrm{old}.
\end{align}
We lift the boosted solution to M-theory by adding an eleventh dimension $y$ in the following S-duality fashion
\begin{align}
\label{eq:Sduality}
ds^2_{11} = e^{-2\phi/3} ds_{10}^2 +  e^{4\phi/3} \left(dy + A_\mu dx^\mu\right)^2.
\end{align}
There is no gauge field $A_\mu$ in our solution and we therefore have,
using $e^{2\phi} =H_1^{-1}$,
\begin{align}
ds_{11}^2 = H_1^{1/3} ds_{10}^2 + H_1^{-2/3} dy^2.
\end{align}
The B field gives rise to a three-form with non-vanishing components
\begin{align}
C_{txy} = \coth\alpha_1  \left(H_1^{-1} - 1\right).
\end{align}
There is no dilaton in 11 dimensions and this is therefore the full solution.

Let us rewrite the boosted part of the metric (the $dt$ and $dx$ components)
before reducing on $x$.  Defining $H_4 \equiv 1+ (1-U)\sinh^2\alpha_4$, we find
\begin{align}
\big(-U\cosh^2\alpha_4 &+ \sinh^2\alpha_4\big) dt^2
                  -  2(1-U) \cosh\alpha_4\sinh\alpha_4 dt dx
                 + \left(-U\sinh^2\alpha_4 + \cosh^2\alpha_4\right)dx^2 \nonumber\\
&= -H_4^{-1} Udt^2 + H_4 \left( dx + \coth\alpha_4 \left(H_4^{-1} -1\right)dt \right)^2.
\end{align}
The total metric can therefore be written as
\begin{align}
ds^2_{11} &= H_1^{1/3}\left\{ H_1^{-1}
   \left[ - H_4^{-1} Udt^2 + H_4 \left( dx + A_t dt \right)^2\right]
   + \sum_{i=1}^4 (du^i)^2 + ds^2_4\right\} + H_1^{-2/3}dy^2\nonumber\\
&= H_1^{-2/3}H_4 \left( dx + A_t dt \right)^2
   + H_1^{-2/3}\left( -H_4^{-1} U dt^2 + dy^2 \right)
   + H_1^{1/3}\left( \sum_{i=1}^4 (du^i)^2 + ds^2_4\right)
\end{align}
with $A_t \equiv \coth\alpha_4 \left(H_4^{-1} -1\right)$.

We can now reduce on $x$ by reading the S-duality transformation
(\ref{eq:Sduality}) backwards. From the factor multiplying the first term we
see that
\begin{align}
e^{2\phi} = H_1^{-1}H_4^{3/2}
\end{align}
and therefore
\begin{align}
ds^2_{10} = H_1^{-1} H_4^{-1/2}\left[-Udt^2 + H_4 dy^2
+ H_1H_4 \left( \sum_{i=1}^4 (du^i)^2 + ds^2_4\right) \right].
\end{align}
The three-form gives back our B field and we also have
a new one-form
\begin{align}
B_{ty} &= \coth\alpha_1 \left(H_1^{-1} -1\right),\\
A_{t} &= \coth\alpha_4 \left(H_4^{-1} -1\right).
\end{align}
This is a two-charge solution and we only need one more.

Before boosting again, let us transform the D0-brane into a D4-brane
by T-dualizing in $u_1, u_2, u_3, u_4$.   The dilaton becomes
\begin{align}
e^{\phi} = \frac{H_1^{-1}H_4^{3/2}}{(H_4^{1/2})^4} = H_1^{-1}H_4^{-1/2},
\end{align}
the B-field is unchanged and the gauge field is simply
\begin{align}
A_{(5)} = \coth \alpha_4 (H_4^{-1} - 1) dt \wedge du^1 \wedge du^2 \wedge du^3 \wedge du^4.
\end{align}
The only part of the metric that changes is in the $u$-directions and we find
\begin{align}
ds^2_{10} =  H_1^{-1} H_4^{-1/2}\left(-Udt^2 + H_4 dy^2
+ H_1\ \sum_{i=1}^4 (du^i)^2 + H_1H_4 ds^2_4 \right).
\end{align}

To add the third charge we lift once more to M-theory, boost and reduce.
The procedure is analogous to what has been done before and the end result
is as given in Section (\ref{sec:adding3charges}).

These U-duality transformations of the solution take place in the string
frame but in the main text of the paper we always use the Einstein frame.
Going to ten-dimensional Einstein frame the metric transforms as
\begin{equation}
g_{\mu\nu}^\textrm{E} = e^{-\phi/2} g_{\mu\nu}^\textrm{string}
\end{equation}
and we get
\begin{equation}
\label{eq:Einsteinmetric}
ds_\textrm{E}^2 = H_1^{-\frac{3}{4}}H_4^{-\frac{3}{8}}H_0^{-\frac{7}{8}}
    \left( -Udt^2  +H_4H_0 dx^2 + H_1H_0 \sum_{i=1}^4 (du^i)^2 + H_1H_4H_0\frac{L^2}{(2\pi)^2} V_{ab} dx^a dx^b  \right).
\end{equation}

\section{Relating the $c$'s and $\bar c$'s}
\label{app:cmap}

To find how  the expansion coefficients of the non-extremal metric
(the $\bar c$'s) are related to the original seeding
solution we recall that the seeding solution has
\begin{align}
-g^\textrm{seed}_{tt} \simeq U = 1 - \frac{c_t}{r},  \qquad
g^\textrm{seed}_{zz} \simeq 1 + \frac{c_z}{r} .
\end{align}
The harmonic functions can then be written as
\begin{equation}
H_a \simeq 1 + \frac{c_t}{r} \sinh^2 \alpha_a
\end{equation}
and we can read the asymptotics off the new metric
\begin{align}
-g_{tt} &= H_1^{-\frac{3}{4}}H_4^{-\frac{3}{8}}H_0^{-\frac{7}{8}}U \\
 &= 1 - \frac{c_t}{r} \left( 1 + \frac{3}{4}\sinh^2 \alpha_1
                                         + \frac{3}{8}\sinh^2 \alpha_4
                                         + \frac{7}{8}\sinh^2 \alpha_0 \right) + \cdots
\end{align}
This shows that
\begin{equation}
\bar c_t = c_t
\left( 1 + \frac{3}{4}\sinh^2 \alpha_1 + \frac{3}{8}\sinh^2 \alpha_4 + \frac{7}{8}\sinh^2 \alpha_0 \right) .
\end{equation}
Similarly we find
\begin{align}
\bar c_x &= c_t
\left(-\frac{3}{4}\sinh^2 \alpha_1 + \frac{5}{8}\sinh^2 \alpha_4 + \frac{1}{8}\sinh^2 \alpha_0 \right),\\
\bar c_u &= c_t
\left(\frac{1}{4}\sinh^2 \alpha_1 - \frac{3}{8}\sinh^2 \alpha_4 + \frac{1}{8}\sinh^2 \alpha_0 \right),\\
\bar c_z &= c_z + c_t
\left(\frac{1}{4}\sinh^2 \alpha_1 + \frac{5}{8}\sinh^2 \alpha_4 + \frac{1}{8}\sinh^2 \alpha_0 \right),
\end{align}
and
\begin{equation}
\bar c_{A_a} =  - c_t \sinh\alpha_a \cosh\alpha_a .
\end{equation}
For the phantom direction $u^0$ the factor in front of $(du^0)^2$
inside the parenthesis in Equation (\ref{eq:Einsteinmetric}) would be $H_1H_4$
(from the harmonic rule of \cite{Peet:2000hn}) and we find
\begin{equation}
\bar c_0 = c_t
\left(\frac{1}{4}\sinh^2 \alpha_1 + \frac{5}{8}\sinh^2 \alpha_4 - \frac{7}{8}\sinh^2 \alpha_0 \right) .
\end{equation}

\section{Electric masses and tensions}\label{app:elec}
In this appendix we examine the electric mass and tensions in
greater detail.

\subsection{Direct calculation}\label{sec:appelecsec1}

In this subsection we will show how to calculate the electric mass
and tensions in 10 and 11 dimensions for general objects composed of
transverse branes, F1-strings etc. obeying the harmonic function
rule, and based on a seeding solution of dimension $d+1$. This will
be done using the method of equivalent sources.

The matter part of the energy momentum tensor, $T^{\mathrm{mat}}$,
consists of a dilatonic part, $T^{\mathrm{dil}}_{\mu\nu}$, and parts
from the gauge field strengths $F_a$, $T^{(F_a)}_{\mu\nu}$:
\begin{equation}\label{eq:appelec1}
    T^{\mathrm{mat}}_{\mu\nu}=T^{\mathrm{dil}}_{\mu\nu}+\sum_{a=1}^{N_{\mathrm{ch}}}T^{(F_a)}_{\mu\nu},
\end{equation}
where $N_{\mathrm{ch}}$ is the number of different charges. The
explicit expressions for the energy-momentum tensors are:
\begin{equation}\label{appelec0.5}
8\pi G_D
T^{\mathrm{dil}}_{\mu\nu}=-\frac{1}{4}g_{\mu\nu}\partial^{\rho}\phi\partial_{\rho}\phi+\frac{1}{2}\partial_{\mu}\phi\partial_{\nu}\phi
\end{equation}
\begin{equation}\label{appelec0.6}
    8\pi G_D T^{(F_a)}_{\mu\nu}=-\frac{1}{2}g_{\mu\nu}\frac{1}{2(p_a+2)!}e^{\ka_a\phi}(F_a)^2+\frac{1}{2(p_a+1)!}e^{\ka_a\phi}(F_a)_\mu^{\phantom{\mu}\rho_1\cdots\rho_{p_a+1}}(F_a)_{\nu\rho_1\cdots\rho_{p_a+1}},
\end{equation}
where $D$ is the total number of dimensions, $p_a$ is the number of
world-volume directions for object $a$, and $\ka_a$ is a number
depending on $p_a$.

To calculate the mass and tension we use the method of equivalent
sources. This means that instead of studying the real metric we will
study a metric with the same asymptotics as our solution, but which
is everywhere Newtonian, i.e. we can split the metric as:
\begin{equation}\label{eq:appelec2}
    g_{\mu\nu}=\eta_{\mu\nu}+h_{\mu\nu},
\end{equation}
where $\eta_{\mu\nu}$ is the Minkowski metric and $h_{\mu\nu}$ is
small so we can ignore second order contributions in $h$.

We now define:
\begin{equation}\label{eq:appelec3}
    S_{\mu\nu}\equiv T_{\mu\nu}-\frac{1}{D-2}g_{\mu\nu}T\simeq
    T_{\mu\nu}-\frac{1}{D-2}\eta_{\mu\nu}T,
\end{equation}
where $D$ is the total number of dimensions and
$T=T^\mu_{\phantom{\mu}\mu}\simeq \eta^{\mu\nu}T_{\nu\mu}$. We can
inverse this as:
\begin{equation}\label{eq:appelec4}
    T_{\mu\nu}\simeq S_{\mu\nu}-\frac{1}{2}\eta_{\mu\nu}S.
\end{equation}
Then Einstein's equation takes the form
\begin{equation}\label{eq:appelec5}
    S^{\mathrm{mat}}_{\mu\nu}=\frac{1}{8\pi G_D}R_{\mu\nu}
\end{equation}
The linearized Ricci tensor is
\begin{equation}\label{eq:appelec6}
    R^{(1)}_{\mu\nu}=-\frac{1}{2}\left(\square h_{\mu\nu}+h_{\la\phantom{\la},\mu\nu}^{\phantom{\la}\la}-h_{\mu\phantom{\la},\nu\la}^{\phantom{\mu}\la}-h_{\nu\phantom{\la},\mu\la}^{\phantom{\nu}\la}\right)
\end{equation}
The gravitational part of the energy-momentum tensor is then defined
from:
\begin{equation}\label{eq:appelec7}
    S_{\mu\nu}^{\mathrm{gr}}\equiv\frac{1}{8\pi G_D}\left(R_{\mu\nu}^{(1)}-R_{\mu\nu}\right)
\end{equation}
Summing the contributions from both the gravitational part and the
matter part gives:
\begin{equation}\label{eq:appelec8}
    S_{\mu\nu}=S_{\mu\nu}^{\mathrm{gr}}+S^{\mathrm{dil}}_{\mu\nu}+\sum_{a=1}^{N_{\mathrm{ch}}}S^{(F_a)}_{\mu\nu}=\frac{1}{8\pi G_D}
    R_{\mu\nu}^{(1)}
\end{equation}
The electric part of the energy momentum tensor will simply be
defined as the part of the tensor that goes to zero when we set the
charges to zero, i.e.:
\begin{equation}\label{eq:appelec8.5}
    S_{\mu\nu}^{\mathrm{el}}\equiv S_{\mu\nu}-S_{\mu\nu}|_{Q_a=0}
\end{equation}
We note that we a priori have contributions from all the parts of
the energy momentum tensor in~\eqref{eq:appelec8}, especially, we
see that both the terms from the gauge fields and the dilatonic term
(the dilaton is constant in the case where the charges are zero) are
completely electric.

Using that all raisings and lowerings can be done with $\eta_{\mu\nu}$, that the
covariant derivatives can be replaced with ordinary, and that
$h_{\mu\nu}$ only depends on $r$ and is diagonal, we get
from~\eqref{eq:appelec4}:
\begin{multline}\label{eq:appelec9}
    T_{\mu\mu}=\frac{1}{16\pi
    G_D}
    \partial_r^2\left(-h_{\mu\mu}-\eta_{\mu\mu}(h_{rr}-\eta^{\alpha\beta}h_{\beta\alpha})\right)
    \\
    =\partial_r^2\left(-h_{\mu\mu}-\eta_{\mu\mu}\left(h_{tt}-\frac{(d-2)}{r^2}h_{\Om\Om}-h_{zz}-\sum_a
    p_{a}h_{u_{(a)}u_{(a)}}\right)\right),
\end{multline}
where $\mu\neq r$, $\Om$ refers to the angular directions, and
$u_{(a)}^i$ are the coordinates for the world-volume for object $a$
which will have a volume denoted by $V_{p_a}$. We see that this is a
boundary term so that we easily can get the masses and tensions
(since $h_{xx}\sim\bar c_{x}/r^{d-3}$):
\begin{equation}\label{eq:appelec10}
    \bar M=\int T_{tt}=\frac{(\prod V_{p_a}) L \Om_{(d-2)}}{16\pi
    G_D}(d-3)\bigg((d-2)\bar c_{\Om}+\bar c_z+\sum_a p_a\bar
    c_{u_{(a)}}\bigg),
\end{equation}
\begin{equation}\label{eq:appelec11}
    L\bar {\cal T}_z=-\int T_{zz}=\frac{(\prod V_{p_a}) L \Om_{(d-2)}}{16\pi
    G_D}(d-3)\bigg((d-2)\bar c_{\Om}+\sum_a p_a\bar
    c_{u_{(a)}}-\bar c_t\bigg),
\end{equation}
\begin{equation}\label{eq:appelec12}
    L_{u_{(a)}}\bar {\cal T}_a=\frac{(\prod V_{p_a}) L \Om_{(d-2)}}{16\pi
    G_D}(d-3)\bigg((d-2)\bar c_{\Om}+\bar c_z+\sum_{a'\neq a} p_{a'}\bar
    c_{u_{(a')}}+(p_a-1)\bar c_{u_{(a)}}-\bar c_t\bigg).
\end{equation}

The asymptotic quantities are determined by the real physical metric
which by the harmonic function rule has components:
\begin{align}
    g_{tt}&=-H\bigg(\prod_a H_{(a)}^{-1}\bigg)U,\quad
    g_{u_{(a)}^iu_{(a)}^i}=H_{(a)}^{-1}H,\nonumber\\
    g_{zz}&=H f_z,\quad g_{\Om\Om}=H f_{\Om},\label{eq:appelec13}
\end{align}
where $U\sim1$, $f_z\sim1$, and $f_{\Om}\sim r^2$ are functions from
the seeding solution, and:
\begin{align}\label{eq:appelec14}
    H\equiv \prod_a H_{(a)}^{\bet_a}, \quad
    \bet_a=\frac{p_a+1}{D-2},
\end{align}
which holds for Dp-branes, F1-strings and NS5-branes in $D=10$
dimensions, and M2- and M5-branes in $D=11$ dimensions. The harmonic
function is assumed to be given by:
\begin{equation}\label{eq:appelec15}
    H_{(a)}=1+(1-U)\sinh^2\al_a.
\end{equation}

The electric part of the mass and tensions are defined
by~\eqref{eq:appelec8.5} as the part that goes to zero when we set
the charges to zero. We can now use the metric to find the electric
parts of $\bar c_t$, $\bar c_z$, etc. in terms of the seeding $c_t$:
\begin{align}
    \bar c_t^{\mathrm{el}}&=\sum_a\frac{D-2-p_a-1}{D-2}\sinh^2 \al_a
    c_t\nonumber\\
    \bar c_{u_{(a)}}^{\mathrm{el}}&=\sum_{a'}\frac{p_{a'}+1}{D-2}\sinh^2 \al_{a'}
    c_t-\sinh^2\al_a c_t\nonumber\\
    \bar c_z^{\mathrm{el}}&=\sum_a\frac{p_a+1}{D-2}\sinh^2 \al_a
    c_t\nonumber\\
    \bar c_\Om^{\mathrm{el}}&=\bar c_z^{\mathrm{el}}\label{appelec16}
\end{align}
Inserting this in~\eqref{eq:appelec10}--\eqref{eq:appelec12} finally
gives:
\begin{align}
    \bar M^{\mathrm{el}}&=\frac{(\prod V_{p_a}) L \Om_{(d-2)}}{16\pi
    G_D}(d-3)c_t\sum_a\sinh^2\al_a\label{appelec17}\\
    L\bar {\cal T}_z^{\mathrm{el}}&=0\label{appelec18}\\
    L_{u_{(a)}}\bar {\cal T}_a^{\mathrm{el}}&=\frac{(\prod V_{p_a}) L \Om_{(d-2)}}{16\pi
    G_D}(d-3)c_t\sinh^2\al_a\label{appelec19}
\end{align}
which, of course, is the same result as~\eqref{eq:Mel}. We note that
since we should look at the asymptotic behavior of our metric, then
by the form of $H_{(a)}$ in~\eqref{eq:appelec15} our electric masses
and tensions split in a sum of contributions from each object $a$.
We also observe that there is no contribution to the electric parts
of the tension in some given direction from objects transverse to
this direction. Especially, the electric part of the tension in the
$z$-direction is zero. We will use this as one of the basic
principle in the next section.

\subsection{Symmetry considerations}\label{sec:appelecsec2}

In this section we will investigate the electric masses and
tensions, but this time based on some simple symmetry consideration
and some physical principles that have been confirmed in the last
subsection. We will follow the analysis in~\cite[section
3.2]{Harmark:2004ws}, but with our definition of the electric mass
and tension.

First, we assume that the electric parts of the
energy-momentum tensor split up into contributions from each of the
objects $a$, as was confirmed in last section, i.e.:
\begin{equation}\label{eq:appelec20}
    T_{\mu\nu}^{\mathrm{el}(a)}=T_{\mu\nu}|_{\forall a'\neq a: Q^{a'}=0}-T_{\mu\nu}|_{\forall a':Q^{a'}=0}.
\end{equation}
This gives rise to the electric parts of the mass and tensions.

We will still use the method of equivalent sources, so that we can
neglect second order contributions in $h_{\mu\nu}$. We then require
boost-invariance for object $a$, i.e.
$T_{tt}^{\textrm{el}(a)}=-T_{u^j_{(a)}u^j_{(a)}}^{\textrm{el}(a)}$.
This can be seen to be fulfilled by the dilatonic and gauge part of
the energy-momentum tensor in~\eqref{appelec0.5}
and~\eqref{appelec0.6} (using that $\phi$ only depends on $r$) and,
actually, also for $R_{\mu\nu}^{(1)}$ using~\eqref{eq:appelec13}
(assuming the symmetry holds for the equivalent metric also)
and~\eqref{eq:appelec6}, and hence for the whole energy-momentum
tensor. After integrating, the boost-invariance implies:
\begin{equation}\label{eq:appelec21}
    M^{\textrm{el}(a)}=L_{u_{(a)}}\bar{\cal
    T}_a^{\textrm{el}(a)}.
\end{equation}

We further require that for $\nu$ a transverse direction
to object $a$ we have (i.e. after integrating the electromagnetic
part of the energy-momentum tensor):
\begin{equation}\label{eq:appelec22}
    L_{\nu}\bar{\cal
    T}_\nu^{\textrm{el}(a)}=-\int
    T_{\nu\nu}^{\textrm{el}(a)}=0,\quad\textrm{$\nu$ transverse to object $a$.}
\end{equation}
This means that we in~\eqref{eq:appelec21} can replace the tension
on the right hand side with the total electric tension in the
$u_{(a)}$-direction:
\begin{equation}\label{eq:appelec23}
    M^{\textrm{el}(a)}=L_{u_{(a)}}\bar{\cal T}_a^{\textrm{el}}.
\end{equation}

Assuming a diagonal energy-momentum tensor and using the method of
equivalent sources we get from~\cite{Harmark:2004ch} that
\begin{equation}\label{eq:appelec24}
    \nabla^2 g_{u_{(a)}^i u_{(a)}^i} = -16\pi
G_{D}\left(T_{u_{(a)}^i u_{(a)}^i} -
\frac{1}{D-2}T^{\rho}_{\phantom{\rho}\rho}\right).
\end{equation}
Taking the non-electric part of this we should set all the charges
to zero. In that case the directions $u_{(a)}^i$ should be
completely flat, and hence the left hand side should be zero, i.e.:
\begin{equation}\label{eq:appelec25}
    T_{u_{(a)}^i
    u_{(a)}^i}^{\textrm{non-el}}=\frac{1}{D-2}\sum_\rho\eta^{\rho\rho}T^{\textrm{non-el}}_{\rho\rho},
\end{equation}
or:
\begin{align}\label{eq:appelec26}
    (D-2)T^{\textrm{non-el}}_{u_{(a)}u_{(a)}}=-T^{\textrm{non-el}}_{tt}+T^{\textrm{non-el}}_{zz}+\sum_{a} p_aT^{\textrm{non-el}}_{u_{(a)}u_{(a)}}+\textrm{overall transverse terms.}
\end{align}
Integrating, using~\eqref{eq:appelec22} and~\eqref{eq:appelec23},
and solving we get:
\begin{equation}\label{eq:appelec27}
    (d-1)(L_{u_{(a)}} {\cal T}_a-M^{\textrm{el}(a)})=\bar
    M-M^{\textrm{el}}+L\bar {\cal T}_z.
\end{equation}
In general these $N_{\mathrm{ch}}$ equations can be solved for the
unknowns $M^{\textrm{el}(a)}$. However, precisely in our
three-charge case with $d-1=3$ the equations have determinant zero.
Adding the three equations gives:
\begin{equation}\label{eq:appelec28}
L_{x}\bar {\cal T}_{x}+L_{u^i}\bar {\cal T}_{u^i}+L_{0}\bar {\cal
T}_{0}=\bar
    M+L\bar {\cal T},
\end{equation}
which exactly is independent of $M^{\textrm{el}}$. From this
equation we, however, obtain the two nice relations:
\begin{align}\label{eq:appelec29}
    \bar n_x+\bar n_0+\bar n_u&=1+\bar n,\\
    r_x+r_0+r_u&=1+r\label{eq:appelec30}
\end{align}
which are indeed obeyed by~\eqref{eq:defbarn}
and~\eqref{eq:defbarna}, and~\eqref{eq:erra}.

For general dimensions and charges we instead get:
\begin{align}\label{eq:appelec31}
    \sum_a \bar n_a&=\frac{N_{\mathrm{ch}}}{d-1}+\frac{N_{\mathrm{ch}}}{d-1}\bar n,\\
    \sum_a r_a&=\left(1-\frac{d-2}{2}\right)\frac{N_{\mathrm{ch}}\mu}{(d-1)\eps}+\frac{3N_{\mathrm{ch}}}{2(d-1)}r,\label{eq:appelec32}
\end{align}
where in the last equation the limit $\al_a\to \infty$ had to be
taken in the same way as in the end of
Section~\ref{sec:finiteentropy}. We see that for $d=4$ the last
relation reduces to:
\begin{equation}\label{appelec33}
\sum_a r_a=\frac{N_{\mathrm{ch}}}{2}r.
\end{equation}
If we use this for the three-charge case we exactly get $r=2$.

Finally, if we use this on near-extremal two-charge case we get:
\begin{equation}\label{appelec}
r_0+r_4=r
\end{equation}
in agreement with~\eqref{maplim1} (we set the last charge to zero).
Using also~\eqref{eq:appelec30} we also conclude that the last
relative tension should be one again in agreement
with~\eqref{maplim1}.

\addcontentsline{toc}{section}{References}

\providecommand{\href}[2]{#2}\begingroup\raggedright\endgroup

\end{document}